\documentclass[%
reprint,
nofootinbib,
amsmath,amssymb,
aps,
]{revtex4-2}

\usepackage{graphicx}
\usepackage{dcolumn}
\usepackage{bm}
\usepackage{hyperref}

\begin{document}
	
	\preprint{APS/123-QED}
	
	\title{Gauge hierarchy from electroweak vacuum metastability}
	
	\author{Justin Khoury}
	\email{jkhoury@sas.upenn.edu}
	\affiliation{%
		Center for Particle Cosmology, Department of Physics and Astronomy, University of Pennsylvania, Philadelphia, PA 19104 
	}%
	
	\author{Thomas Steingasser}
	\email{Thomas.Steingasser@physik.lmu.de}
	\affiliation{
		Arnold Sommerfeld Center for Theoretical Physics, Ludwig-Maximilians-Universit\"at, Theresienstra\ss e 37, 80333 Munich
	}%

	\date{\today}
	
	\begin{abstract}
		We consider the possibility that the gauge hierarchy is a byproduct of the metastability of the electroweak vacuum, {\it i.e.}, that whatever mechanism is responsible for the latter also sets the running Higgs mass to a value smaller than its natural value by many orders of magnitude. We find that the metastability of the electroweak vacuum, together with the requirement that such a non-trivial vacuum exists, requires the Higgs mass to be smaller than the instability scale by around one order of magnitude. While this bound is quite weak in the Standard Model (SM), as the instability scale is~$\sim 10^{11}$~GeV, simple and well-motivated extensions of the SM significantly tighten the bound by lowering the instability scale. We first consider the effect of right-handed neutrinos in the~$\nu$MSM with approximate~$B-\Tilde{L}$ symmetry, which allows for masses of order~TeV for the right-handed neutrinos and~$\mathcal{O}(1)$ Yukawa couplings. We find that right-handed neutrinos cannot by themselves fully explain the gauge hierarchy, as the tightest upper bound compatible with current experimental constraints is~$\sim 10^8$~GeV. As we demonstrate on the example of the minimal~SU(4)/Sp(4) composite Higgs model, this bound can be lowered significantly through the interplay of the neutrinos and a dimension-six operator. In this scenario, the bound can be brought down considerably, with the smallest value accessible by our perturbative treatment being of order~$\simeq 10$~TeV, and consistently several orders of magnitude below its natural value. While this is insufficient to fully solve the gauge hierarchy problem, our results imply that, assuming the SM symmetry breaking pattern, small running Higgs masses are a universal property of theories giving rise to metastability, suggesting a common origin of the two underlying fine-tunings and providing a strong constraint on any attempt to explain metastability.
	\end{abstract}
	
	\maketitle
	

	\section{Introduction}
	The principles of \textit{naturalness} and symmetry have guided the development of fundamental physics over the last fifty years. However, in the light of our universe's apparent fine-tuning, it appears increasingly likely that particle physics has entered a \textit{post-naturalness} era~\cite{Giudice:2017pzm}. One of the most important observations motivating this perspective are the measured values of the Higgs and top quark mass,~$M_h=125.1$~GeV and~$M_t=172.4$~GeV, which imply that the Standard Model (SM) couplings remain perturbative until energies exceeding even the Planck scale. This, together with the absence of new physics in flavor, precision and LHC experiments, points to the SM being valid up to very high energies, and potentially even the Planck scale. 
	
	An obvious drawback of such a \textit{grand desert} above the electroweak scale is that the Higgs mass requires fine-tuning to cancel large radiative corrections, which cannot be explained by (technical) naturalness alone~\cite{tHooft:1979rat}, leading to the so-called \textit{hierarchy problem}. Furthermore, it opens up the question why the laws of nature appear to include two fundamental scales, set by the Planck and the Higgs mass, which are separated from one another by~$17$ orders of magnitude, which is known as the \textit{Higgs naturalness} or \textit{gauge hierarchy problem}. 
	
	As we argue throughout this article, this last question might in parts be answered by considering yet another apparent fine-tuning of the Higgs sector. The masses of both the Higgs and the top quark lie in an extraordinarily small window corresponding to a metastable electroweak vacuum~\cite{Frampton:1976kf,Sher:1988mj,Casas:1994qy,Espinosa:1995se,Isidori:2001bm,Espinosa:2007qp,Ellis:2009tp,Degrassi:2012ry,Buttazzo:2013uya,Lalak:2014qua,Andreassen:2014gha,Branchina:2014rva,Bednyakov:2015sca,Iacobellis:2016eof,AFS}. When extrapolated to high energies, the Higgs quartic coupling becomes negative at the instability scale~$\mu_I \sim 10^{11}$~GeV and remains small, which allows our vacuum to decay through bubble nucleation. Using the most recent global averages given in~\cite{Zyla:2020zbs,Huang:2020hdv}, its lifetime, defined as the characteristic time to form a bubble a true vacuum within our past light-cone, is found to be at~$1\sigma$\footnote{We take into account the correlated errors in the top Yukawa, Higgs quartic and strong gauge coupling as given in~\cite{Huang:2020hdv}. An extensive discussion of the lifetime's sensitivity to other parameters can be found in~\cite{Andreassen:2014gha}.}
	\begin{equation}
		\tau_{\rm EW} \sim 10^{983^{+1410}_{-430}}~{\rm years}\,.
		\label{lifetime intro}
	\end{equation}
	This hinges on a delicate cancellation between the exponentially small decay rate per unit volume of the vacuum,~$\frac{\Gamma}{V}\sim \exp\left(- \frac{8\pi^2}{3|\lambda(\mu_\star)|}\right)$, which in the pure Standard Model is exquisitely sensitive to the quartic coupling at the scale~$\mu_\star \sim  10^{17}~{\rm GeV}$ where the latter achieves a minimum, and the exponentially large space-time volume of the observable universe. In other words, metastability requires a remarkable conspiracy between the cosmological constant and the couplings of the Standard Model.  
	
	A way forward is suggested by a common theme underlying many fine-tunings, namely that they are problems of near-criticality. The metastability of the electroweak vacuum requires that the decay rate per space-time volume be relatively close to the critical range for the percolation phase transition of bubble nucleation~\cite{Guth:1982pn}. The gauge hierarchy problem can be interpreted as the Higgs having a nearly vanishing mass relative to the fundamental scale, close to the phase transition between broken/unbroken electroweak symmetry~\cite{Giudice:2006sn}.\footnote{However, even with a positive Higgs mass-squared, the electroweak phase would still be spontaneously broken at the QCD scale by the Higgs coupling to the quark condensate. See~\cite{ArkaniHamed:2005yv,Samuel:1999am} for discussions of this other phase of the SM. Thus one must not only explain why the Higgs mass is small, but also why the Higgs mass-squared is negative.} 
	
	The main point of this article is that two of these examples --- the metastability of the electroweak vacuum and the smallness of the (running) Higgs mass compared to the Planck mass --- are in fact not entirely independent, as a small Higgs mass is a necessary condition for metastability. In other words, {\it any} explanation of metastability also offers a path towards a solution to the Higgs naturalness problem.

	\subsection{Gauge hierarchy from metastability}
	Our central result is that that the metastability of the electroweak vacuum, together with the very requirement that such a non-trivial vacuum exists, implies an upper bound on the Higgs mass. In other words, metastability necessarily implies a hierarchy between the running Higgs mass and its natural value, whose precise extent depends on the lifetime.
	
	We show this in three steps:
	\begin{enumerate}
		\item Under our assumptions, the running Higgs mass is bounded from above through the instability scale, $m_h^2 < \left\vert\beta_\lambda (\mu_I)\right\vert (\dots) \mu_I^2$.
		\item The instability scale lies below the so-called \textit{instanton scale}, $\mu_I^2 \ll \mu_S^2$, which we introduce in the next section.
		\item The instanton scale lies below the natural value for the running Higgs mass, $\mu_S^2 \lesssim m_{h, \text{nat}}^2$.
	\end{enumerate}
	We furthermore show that through the effects of well-motivated SM extensions it is possible to obtain a bound just two orders of magnitude above the observed value within the limitations of our perturbative treatment, and potentially even lower if a more elaborate analysis were to be performed.
	
	The relation between Higgs mass parameter and instability scale was first pointed out in the context of the SM in~\cite{Buttazzo:2013uya}, whose authors obtained the following upper bound:\footnote{In Sec.~\ref{higgs mass upper bound} we review its derivation and derive a more precise version of the bound.}
	\begin{equation}
		m^2_h  <  \left\vert\beta_\lambda (\mu_I)\right\vert {\rm e}^{-3/2} \mu_I^2\,. 
		\label{minst intro}
	\end{equation}
	Thus, once the instability scale~$\mu_I$ has been determined, the condition for the existence of a false vacuum constrains the Higgs mass by about one more order of magnitude. 
	
	Until recently, the usefulness of this bound has widely been dismissed, in particular by the authors of~\cite{Buttazzo:2013uya} themselves, as it appears to suffer from two important problems. The first of them is that it simply shifts the question to why the instability scale lies so far below the natural value for the Higgs mass and, more importantly, why it exists in the first place. From our perspective, the existence of the instability scale is given, as it is necessary to allow the vacuum to decay. To achieve metastability with a relatively short lifetime, the quartic coupling must become negative at the instability scale~$\mu_I$, and then continue to fall off until it reaches a sufficiently negative value~$\lambda (\mu_S)$ to obtain the desired lifetime, with~$\mu_S$ being the \textit{instanton scale} defined in Sec.~\ref{EW vacuum decay}. Since~$\lambda$ depends only logarithmically on the scale, this running stretches over several orders of magnitude. Furthermore, as we argue in section II, the properties of the decay rate imply a hierarchy between $\mu_S$ and the natural value for $m_h^2$.
	
	The second reason why one might question the usefulness of~\eqref{minst intro} is that it only requires the Higgs mass to lie slightly below the instability scale, which, assuming continued validity of the SM, can be found at roughly~$10^{11}$~GeV, leaving unexplained an additional~8~orders of magnitude to reach the electroweak scale. It is now, however, important to observe that the precise extent of this remaining gap is sensitive to beyond-the-SM physics. In the context of the usefulness of inequality \eqref{minst intro}, this has recently been discussed in~\cite{Giudice:2021viw}, whose authors argue that the interplay of a dimension-six operator and additional fermions can significantly lower the instability scale, and thus strengthen the bound \eqref{minst intro}. In section II, we discuss how such SM extensions effect the vacuum's decay rate, before providing explicit numerical results for the SM with gravity as well as two well-motivated Standard Model extensions, namely the minimal~SU(4)/Sp(4) composite Higgs model, which serves as an example for the impact of a dimension-6 term, and the~$\nu$MSM with approximate~$B-\Tilde{L}$ symmetry, whose right-handed neutrinos serve as example for additional fermion singlets. Besides lowering the instability scale and thus our bound on the Higgs mass, we find that the influence of the additional fermions also shortens the vacuum's lifetime, to the extent that, \textit{e.g.}, the Page time can be achieved. We provide accurate predictions of the vacuum's lifetime as a function of the parameters characterizing the SM extensions of interest. As an important side result, we not only update existing stability bounds by combining for the first time all relevant NLO corrections, both from gravity and functional determinants, at up to 3-loop accuracy, but also provide a translation of any prediction of the vacuum's lifetime to parameters of the considered SM extensions. 
	
	Before discussing in more detail the SM extensions considered here, it is worth emphasizing that our upper bound~\eqref{minst intro} relies on two assumptions: i)~metastability of the electroweak vacuum; ii)~the existence of such a non-trivial vacuum, {\it i.e.}, a negative Higgs mass-squared. 
	
	\subsection{Achieving metastability}
	One framework in which metastability may naturally arise is the landscape of string theory, together with the mechanism of eternal inflation~\cite{Steinhardt:1982kg,Vilenkin:1983xq,Linde:1986fc,Linde:1986fd,Starobinsky:1986fx} for dynamically populating its vacua.  
	The \textit{principle of mediocrity}, generally assumed in this context ({\it e.g.},~\cite{Garriga:2005av}), entails that we live at asymptotically late times in the unfolding of eternal inflation, when the relative occupational probabilities of different vacua have settled to a near-equilibrium distribution. 
	
	However, a more interesting possibility for our purposes is based on the idea that we exist during the approach to equilibrium, {\it i.e.}, at times much earlier than the exponentially-long mixing time for the landscape. In this case, a vacuum like ours should be likely not because it is typical according to a quasi-stationary distribution, but rather because it has the right properties to be accessed early on in the evolution~\cite{Denef:2017cxt}. This perspective offers a dynamical selection mechanism for vacua based on search optimization~\cite{Khoury:2019yoo,Khoury:2019ajl,Kartvelishvili:2020thd,Khoury:2021grg}: vacua that are easily accessed reside in optimal regions where the search algorithm defined by local landscape dynamics is efficient. This idea was formalized with the definition of an accessibility or early-time measure in~\cite{Khoury:2019ajl,Khoury:2021grg}. Importantly, optimal regions of the landscape display non-equilibrium critical phenomena, in the sense that their vacuum dynamics are tuned at {\it dynamical criticality}. This suggests a deep connection between the near-criticality of our universe and non-equilibrium phase transitions on the landscape.  
	
	A key prediction is that optimal regions are characterized by vacua that are relatively short-lived, with lifetimes of order their de Sitter Page time.\footnote{It is intriguing that the de Sitter Page time has also emerged recently in a completely different context, as the quantum break time of de Sitter space~\cite{QBT}.} For our vacuum, this optimal lifetime is 
	\begin{equation}
		\tau_{\rm Page} \sim \frac{M_{\rm Pl}^2}{H^3_0} \simeq 10^{130}~{\rm years}\,.
		\label{tau Page intro}
	\end{equation}
	While this lies around~$850$ orders of magnitude below the central value of the SM, it should be clear that the latter is highly sensitive to beyond-the-SM physics. Indeed, as we will show, well-motivated extensions of the SM, such as right-handed neutrinos, can very well shorten the vacuum lifetime down to the Page time. 
	
	Importantly, the early-time approach to eternal inflation offers a {\it raison d'\^{e}tre} for the conspiracy underlying Higgs metastability. That being said, while we are primarily motivated by non-equilibrium eternal inflationary dynamics, it is worth emphasizing once again that our analysis is logically independent of this proposal. The results below pertain more generally to {\it any} theoretical framework which predicts that our vacuum should be metastable. 
	
	One other example for such a framework is described in~\cite{Giudice:2021viw}. These authors propose that the parameters of the Higgs potential are functions of an additional scalar field, the \textit{apeiron}, and provide a dynamical mechanism based on quantum first-order phase transitions in the early universe favoring values of the field corresponding to a near-critical Higgs potential. For the quartic coupling~$\lambda$, this translates to the selection of an RG trajectory close to the transition from a stable vacuum to a potential with both a false vacuum at the electroweak scale and a true vacuum at some higher energy, albeit the existence of the latter requires some UV completion to stabilize it. As we argue in Sec.~\ref{higgs mass upper bound}, these are the necessary conditions for metastability, while the favored RG trajectory of~$\lambda$ amounts to the prediction of an ideal lifetime. 
	
	What makes this approach particularly interesting from the perspective of this article is that it allows for a prediction of the Higgs mass parameter and vacuum expectation value. Following the reasoning of~\cite{Geller:2018xvz,Giudice:2021viw}, the probability distributions for these quantities are strongly peaked around values near the instability scale. In Appendix~\ref{upper bound m}, we show that this prediction corresponds to the saturation of our bound, making it an important special case of our universally-applicable result.

	\section{Electroweak vacuum decay}
	\label{EW vacuum decay}
	
	We begin with a brief review of Higgs metastability and how it is effected by the SM extensions of interest for our later reasoning, in particular a $H^6$-operator. The electroweak vacuum corresponds to the minimum of the Higgs potential, which at tree level is given by
	\begin{equation}
		V(\mathbf{H})= - \frac{1}{2}m_h^2 |\mathbf{H}|^2 + \lambda  |\mathbf{H}|^4\,,
		\label{Higgstree}
	\end{equation}
	with~$\mathbf{H}$ denoting the standard Higgs doublet. The Higgs boson arises from fluctuations around the minimum of this potential, which can in unitary gauge be parametrized as
	\begin{equation}
		\mathbf{H}= \left(0,\frac{H}{\sqrt{2}}\right)\,; \qquad H = v +h \,,
	\end{equation}
	where~$v = \frac{m_h}{\sqrt{2\lambda}}$ corresponds to the potential's minimum, and~$m^2 = 2 \lambda v^2=m_h^2$ to the tree-level mass of the fluctuation~$h$. In a similar manner, the masses of the fermions as well as the massive gauge bosons are proportional to~$v$, as they are generated through the coupling of these particles to the Higgs field. Thus, most particles of the SM can be found at energies somewhere below~$v$. That is,~$v$, or equivalently~$m_h$, sets the electroweak scale.
	
	In a purely classical theory, the configuration~$H=v$ represents the theory's unique, stable vacuum up to SU(2)-transformations of the full Higgs-doublet. This is, however, not necessarily true once the scale dependence of the parameters, due to quantum effects, is taken into account. Doing so gives rise to the RG-improved effective potential, which, for energies well above the electroweak scale, takes the simple form
	\begin{equation}
		V_{\text{eff}}(h)\simeq  \frac{1}{4} \lambda_{\text{eff}} (h) h^4\,,
		\label{Veff4}
	\end{equation}
	where~$\lambda_{\text{eff}} (h)$ combines the running quartic coupling, its loop corrections and the Higgs' wave function renormalization factors.
	
	Depending on the precise values of the couplings, it is possible for~$\lambda$ to turn negative at high energies~\cite{Sher:1988mj}. The data compiled at the LHC indicates that this might indeed be the case, as the measured values of the quartic and Yukawa couplings lie just shy of the critical value beyond which~$\lambda$ would remain positive at all scales. For the most accurate values of the couplings available,~$\lambda$ turns negative at the instability scale~$\mu_I \simeq 10^{11}$~GeV, which signals metastability --- the electroweak vacuum can decay through nucleation of a bubble of large Higgs field within a small region of space, which then expands indefinitely.
	
	The lifetime of the electroweak vacuum is determined by the bubble nucleation rate per unit volume,~$\Gamma/V$. Its dominant contribution arises from the so-called bounce solution, which is the instanton describing the tunneling from the false vacuum into the region of negative~$\lambda$, as famously shown in~\cite{Coleman,Callan:1977pt,Coleman:1980aw} and built upon in~\cite{Fubini,Lipatov}. Once the rate per unit volume is known, it is straightforward to obtain the lifetime~$\tau_{\text{EW}}$ of the vacuum, {\it i.e.}, the time at which the probability that a bubble has formed within an observer's past light-cone becomes unity. Assuming vacuum energy dominance, it is given by
	\begin{equation}
		\tau_{\text{EW}} =\frac{3 H^3_0}{4 \pi} \bigg(\frac{\Gamma}{V} \bigg)^{-1}\,, 
		\label{ltew}
	\end{equation}
	where~$H_0$ is the Hubble constant. It is also common in the literature to define the lifetime as the characteristic time scale $\tau_{\text{EW}} \sim (\frac{\Gamma}{V})^{- \frac{1}{4}}$, as is, e.g., the case in \cite{AFS}\footnote{The difference in our lifetime is also partially due to an increase in the world average of the top mass since \cite{AFS} was published.}.
	
	An additional technical hurdle for a potential of the form~\eqref{Veff4} arises from its classical scale invariance. As a result, there exists not just one bounce configuration, but a one-parameter family of solutions, 
	\begin{equation}
		H_R(r)= \frac{2\sqrt{2}}{\sqrt{-\lambda}} \frac{R}{R^2+r^2}\,, 
		\label{bounce}
	\end{equation}
	whose parameter~$R$ is the size of the bubble of true vacuum, and~$r$ is the four-dimensional Euclidean radius. This degeneracy manifests itself in the form of a non-trivial zero mode, the dilatation mode, which on its own would result in a divergent decay rate. An extensive discussion of the SM lifetime, including a clear solution to the scale invariance problem, was recently given in~\cite{AFS}. The problem is naturally solved by the RG running of~$\lambda$, which breaks the scale invariance radiatively, resulting in the decay rate being dominated by the bounce of size~$R= \mu_\star^{-1}$, where~$\mu_\star$ satisfies
	\begin{equation}
		\beta_\lambda (\mu_\star)=0\,. 
		\label{mu* def}
	\end{equation}
	Intuitively, the decay rate is dominated by the configuration associated with the scale at which~$\lambda$ is smallest. 
	
	Before performing the integral over the dilatation mode, the decay rate at NLO takes the form
	\begin{equation}
		\frac{\Gamma}{V}=  \int \frac{\text{d}R}{R^5} \ {\rm e}^{- \frac{8 \pi^2}{3  \left\vert\lambda \left(R^{-1}\right)\right\vert}} D \big(R^{-1}\big)\,,
		\label{DecayR}
	\end{equation}
	where~$D$ summarizes all but the gravitational corrections at NLO, which we give in the Appendix. The integral over~$R$ can be performed explicitly by resumming loop corrections to the Euclidean action, leading to the result given in~\cite{AFS}.\footnote{Note that we evaluate all couplings at the scale~$R^{-1}$, while in~\cite{AFS} they are evaluated at~$\mu_\star$, which is more convenient if one wishes to perform the integral exactly. These authors show that setting the scale equal to~$\mu_\star$ and demanding scale invariance of the nucleation rate induces a series of quantum corrections, allowing them to perform the integral order by order and finally resumming all terms obtained in this way. In order to obtain~\eqref{GravInt}, all one has to do is perform the sum over all the corrections right away in the exponent.} 
	
	As the scale~$\mu_\star$ lies quite close to the Planck mass, high-precision predictions of the lifetime require an analysis of gravitational corrections. Derivations of the leading-order contributions arising from gravity can be found in, {\it e.g.},~\cite{Jose,Gra,Grav}. This complicates the above prescription as it introduces terms in the Euclidean action that explicitly break scale invariance. The general influence of such terms on the decay rate has been discussed with a special focus on the interplay of gravitational corrections with RG running in~\cite{Jose}. In general, introducing an explicit breaking of scale invariance affects the Euclidean action not only directly via the additional term, but also indirectly by changing the scale of the dominant bounce.  
	
	One way to understand this is to reconsider~\eqref{DecayR}, supplemented with a gravitational term for concreteness. We assume that the latter is subdominant, so that the saddle point of the path integral can be approximated to leading order by the family of bounces~\eqref{bounce}. Including the leading-order gravitational correction to the Euclidean action given in, {\it e.g.},~\cite{Grav} leads to a modified version of~\eqref{DecayR},
	\begin{equation}
		\frac{\Gamma}{V} = \int \frac{\text{d}R}{R^5} \ {\rm e}^{- \frac{8 \pi^2}{3 \left\vert\lambda \left(R^{-1}\right)\right\vert}-\frac{256 \pi^3}{45 \lambda^2 \left(R^{-1}\right)} \frac{1}{\left(R M_{\text{Pl}}\right)^2} } D \big(R^{-1}\big)\,. 
		\label{GravInt}
	\end{equation}
	The technique used in~\cite{AFS} to perform the~$R$-integral is now spoiled by the gravitational term, so that we must be content with a saddle-point approximation\footnote{In the following analytical discussions, we treat the $R^{-5}$-prefactor as effectively $O(1)$ when performing the saddle-point approximation, as its dependence on $R$ is negligible compared to the $e^{\frac{1}{\lambda}}$ of the euclidean action. We nevertheless included it in our numerical calculations, confirming that this approximation is well-justified.}:
	\begin{equation}
		\frac{\Gamma}{V} \simeq {\rm e}^{-S_{\text{E}}\left(\lambda (\mu_S),\mu_S\right)} \sqrt{\frac{2 \pi}{  \frac{\text{d}^2}{\text{d} \ln \mu^2}S_{\text{E}}\left(\lambda (\mu_S),\mu_S\right)}} \,\mu_S^4 D (\mu_S)\,, \label{ratefull}
	\end{equation}
	where the Euclidean action is
	\begin{equation}
		S_{\text{E}}\left(\lambda (\mu_S),\mu_S\right) = \frac{8 \pi^2}{3 |\lambda (\mu_S)|}+\frac{256 \pi^3}{45 \lambda^2 (\mu_S)} \frac{\mu_S^2}{M_{\text{Pl}}^2} \,.
		\label{SE defn}
	\end{equation}
	All couplings, as well as the corrections summarized in~$D$, are evaluated at the saddle point~$R^{-1} =\mu_S$ minimizing the Euclidean action, which is the solution of
	\begin{equation}
		\beta_\lambda (\mu_S) \bigg( \frac{64 \pi}{15} \frac{\mu_S^2}{M_{\text{Pl}}^2} - \lambda (\mu_S) \bigg) = \lambda (\mu_S)  \frac{64 \pi}{15} \frac{\mu_S^2}{M_{\text{Pl}}^2}\,.
		\label{saddle}
	\end{equation}
	This result agrees up to a factor of~$\mathcal{O}(1)$ with the rate given in~\cite{Grav}. It is straightforward to prove that~$\mu_S < \mu_*$ in general, with approximate equality~$\mu_S \simeq \mu_\star$ whenever~$\mu_\star \ll M_{\text{Pl}}$.  
	
	As the left-hand side is suppressed by~$\beta_\lambda$, it can be expected that the effect of the gravitational term, which is to lower~$\mu_S$ relative to~$\mu_*$, becomes dominant already for values of~$\mu_S$ one order of magnitude below~$M_{\text{Pl}}$, leading to sub-Planckian values of~$\mu_S$ even for values of~$\mu_*$ multiple orders of magnitude above~$M_{\text{Pl}}$. As this would clearly remain true if one were to replace the Planck mass by any cutoff scale~$\Lambda$, it can in general be expected that any new physics stabilizing the Standard Model below~$\mu_*$ has a strong impact on the decay rate even if the direct correction to the Euclidean action is suppressed by the naturally-small factor~$\frac{\mu_S^2}{\Lambda^2}$. Therefore, we will be careful in what follows to distinguish between these two scales. 
	
	This effect also arises naturally in SM extensions that lead to an extended Higgs potential, most importantly (partially) composite Higgs models and models with a second, heavy Higgs doublet like the ($\nu$)MSSM. After integrating out all heavier degrees of freedom, the most general Higgs potential takes the form
	\begin{equation}
		V_{\text{full}}(H) = - \frac{m_h^2}{4} H^2  + \frac{\lambda}{4} H^4  + \frac{C_6}{\Lambda^2} H^6  + \ldots  ,
		\label{VSMEFT}
	\end{equation}
	where the values of the Wilson coefficients at the matching scale~$\Lambda$ can be obtained by standard techniques. Through power counting and anticipating that $\mu_S$ tends to be smaller than $\Lambda$, it is clear that for a majority of generic potentials the dimension-six term will be the most important, so that we focus on its effect throughout the remainder of this article.
	
	At lower energy, the coefficients~$\{C_{2n}\}_{n>2}$ are obtained by integrating their beta functions. In principle, the dimension-six correction induces additional terms in the SM couplings' beta functions. However, these are suppressed by a factor of~$\frac{m_h^2}{\Lambda^2}$~\cite{SMEFTBeta}, which, as we are about to show, is necessarily~$\ll 1$ for metastable vacua. Hence we can safely neglect these corrections in the beta functions. The main impact of the dimension-six operator will be on the instanton scale~$\mu_S$ and corresponding decay rate.  
	
	We proceed to calculate the vacuum decay rate, including the dimension-six correction to the potential
	\begin{equation}
		\Delta V_{\rm eff} = \frac{C_6}{\Lambda^2} H^6.
		\label{DeltaVCH}
	\end{equation}
	As such, the impact of beyond-the-SM physics is treated perturbatively, neglecting higher-order terms in the full potential~\eqref{VSMEFT}. This approximation is justified provided that
	\begin{equation}
		\left\vert\frac{\Delta V_{\rm eff}}{V_{\rm eff}}\right\vert=  \frac{4 C_6 }{|\lambda |}  \cdot \frac{H^2}{\Lambda^2} \ll 1\,.  
		\label{Del V over V gen}
	\end{equation}
	On the bounce solution, which to leading order is given by~\eqref{bounce}, clearly~$H_R(r)$ is maximal at the origin:~$H_R(r) \leq  H_R(0) \simeq 2\sqrt{2} \mu_S/\sqrt{|\lambda(\mu_S)|}$, where we have focused on the dominant bounce with~$R = \mu_S^{-1}$. Combining with~\eqref{Del V over V gen}, we are led to define an expansion parameter 
	\begin{equation}
		\epsilon \equiv  \frac{32 C_6}{\lambda^2}  \cdot \frac{\mu_S^2}{\Lambda^2}.
		\label{eps def}
	\end{equation}
	For concreteness, we define the regime of validity of our perturbative treatment as~$\epsilon \leq 1$. Given the relation between the lifetime and~$\lambda$, our perturbative treatment is expected to break down either for small values of~$\Lambda$, or for sufficiently long-lived vacua. 
	
	Including the dimension-six operator~\eqref{DeltaVCH} to the Higgs potential yields a correction to the bounce Euclidean action~\eqref{SE defn} given by
	\begin{equation}
		\Delta S_{\rm E}=  \frac{128 \pi^2}{5 |\lambda|^3} \frac{C_6}{R^2 \Lambda^2}\,.
		\label{DeltaSCH}
	\end{equation}
	Thus the decay rate~\eqref{GravInt} becomes, to leading order,
	\begin{widetext}
		\begin{equation}
			\frac{\Gamma}{V} = \int \frac{\text{d}R}{R^5} \exp \bigg(- \frac{8 \pi^2}{3 \left\vert\lambda \left(R^{-1}\right)\right\vert} -\frac{128 \pi^2  (R^{-1})}{5 |\lambda (R^{-1}) |^3}\frac{C_6}{(R \Lambda)^2} -\frac{256 \pi^3}{45 \lambda^2 \left(R^{-1}\right)} \frac{1}{\left(R M_{\text{Pl}}\right)^2} \bigg) \cdot D \big(R^{-1}\big)\,. 
		\end{equation}
	\end{widetext}
	By analogy with the gravitational correction, it is easy to see that the new term has two effects. Firstly, it increases the value of the Euclidean action, thereby stabilizing the vacuum. Clearly the strength of this effect increases with~$\frac{C_6}{\Lambda^2}$, {\it i.e.}, it is most significant either for large~$C_6$ or small~$\Lambda$. Secondly, it shifts the instanton scale~$\mu_S$ further away from~$\mu_\star$. For~$\Lambda \lesssim \mu_\star$, in particular,~$\mu_S$ is shifted to scales below or close to~$\Lambda$.  
	
	This can be quantified in parallel to the treatment of the gravitational term. As before we perform the~$R$-integral in the saddle point approximation, with the scale~$\mu_S$ of the dominant bounce satisfying the modified equation:
	\begin{widetext}
		\begin{equation}
			\beta_\lambda (\mu_S) \bigg( \frac{64 \pi}{15} \frac{\mu_S^2}{M_{\text{Pl}}^2}  - \lambda (\mu_S) - \frac{144}{5} \frac{C_6}{\lambda (\mu_S)} \frac{\mu_S^2}{\Lambda^2} \bigg) +  \beta_{C_6} (\mu_S) \frac{48}{5} \frac{\mu_S^2}{\Lambda^2} = \frac{64 \pi}{15} \lambda (\mu_S) \frac{\mu_S^2}{M_{\text{Pl}}^2} - \frac{96}{5} C_6 (\mu_S) \frac{\mu_S^2}{\Lambda^2}\, .
			\label{beta6}
		\end{equation}
	\end{widetext}
	The decay rate in this approximation is once again given by~\eqref{ratefull}, with the Euclidean action now of the form
	\begin{equation}
		S_{\text{E}}= \frac{8 \pi^2}{3 |\lambda (\mu_S)|}  + \frac{128 \pi^2 C_6 (\mu_S)}{5 |\lambda (\mu_S) |^3} \frac{\mu_S^2}{\Lambda^2}+\frac{256 \pi^3}{45 \lambda^2 (\mu_S)} \frac{\mu_S^2}{M_{\text{Pl}}^2}  \,.
		\label{saddle6}
	\end{equation}

	\section{Upper bound on the running Higgs mass from metastability}
	\label{higgs mass upper bound}
	
	A key point of our analysis is the observation that the metastability of the electroweak vacuum, together with the very requirement that such a non-trivial vacuum exists, implies an upper bound on the Higgs mass. This connection was first pointed out in the context of the SM in~\cite{Buttazzo:2013uya}, though its usefulness was quickly dismissed as the resulting bound lies several orders of magnitude above the observed value. 
	
	At first sight, how the vacuum lifetime can translate to a bound on the Higgs mass is somewhat non-trivial. The lifetime is highly sensitive to the quartic coupling, but it has no explicit dependence on the vacuum expectation value or, equivalently, the Higgs mass. The additional necessary ingredient is the demand of a non-trivial electroweak vacuum. This requires the existence of a minimum of the effective potential at field values below the instability scale~$\mu_I$, where the quartic coupling becomes negative. 
	
	To see this, let us follow~\cite{Buttazzo:2013uya} and minimize the effective potential, including the logarithmic running of~$\lambda$ but neglecting the running of~$m_h$. Demanding that~$\frac{\text{d}}{\text{d}h}V_{\text{eff}}(h)=0$, we find
	\begin{equation}
		m^2_h = \left(2\lambda(v) + \frac{\beta_\lambda(v)}{2} \right)v^2  \,.
	\end{equation}
	At scales somewhat larger than the instability scale, both quadratic and quartic terms appear with a negative sign, so that no extremum can occur. Thus, since we are interested in an upper bound for~$m_h$, we can focus on values of~$v$ near~$\mu_I$, such that~$\lambda (v) \simeq \beta_\lambda (\mu_I) \ln \frac{v}{\mu_I}$. This gives
	\begin{equation}
		m^2_h \simeq  \bigg(2 \ln \frac{v}{\mu_I}  + \frac{1}{2} \bigg)\beta_\lambda (\mu_I)\,v^2 \,. 
		\label{m^2=}
	\end{equation}
	If understood as a function of~$v$, the right-hand side is bounded from above since~$\beta_\lambda (\mu_I) <0$, which implies that a solution can only exist for sufficiently small~$m_h$,
	\begin{equation}
		m^2_h  \lesssim  \left\vert\beta_\lambda (\mu_I)\right\vert {\rm e}^{-3/2} \mu_I^2\,. 
		\label{minst}
	\end{equation}
	Thus, once the instability scale has been determined, the condition for the existence of a false vacuum constrains the Higgs mass by about one more order of magnitude. 
	
	A more precise version of this bound can be obtained by including the non-logarithmic one-loop corrections to~$\lambda$, following the power counting scheme developed in~\cite{Consistent}. As shown in the Appendix~\ref{upper bound m}, doing so replaces~\eqref{minst} by
	\begin{equation}
		m_h^2 \lesssim  \left\vert\beta_\lambda (\mu_I)\right\vert \exp\left(-\frac{3}{2} -2 \frac{\lambda_1 (\mu_I)}{\beta_\lambda (\mu_I)}\right)\mu_I^2\,.
		\label{Ineq}
	\end{equation}
	This of course reduces to~\eqref{minst} if~$\lambda_1$ is neglected, as assumed in~\cite{Buttazzo:2013uya}.
	\begin{figure}[h]
		\centering
		\includegraphics[width=8.6cm]{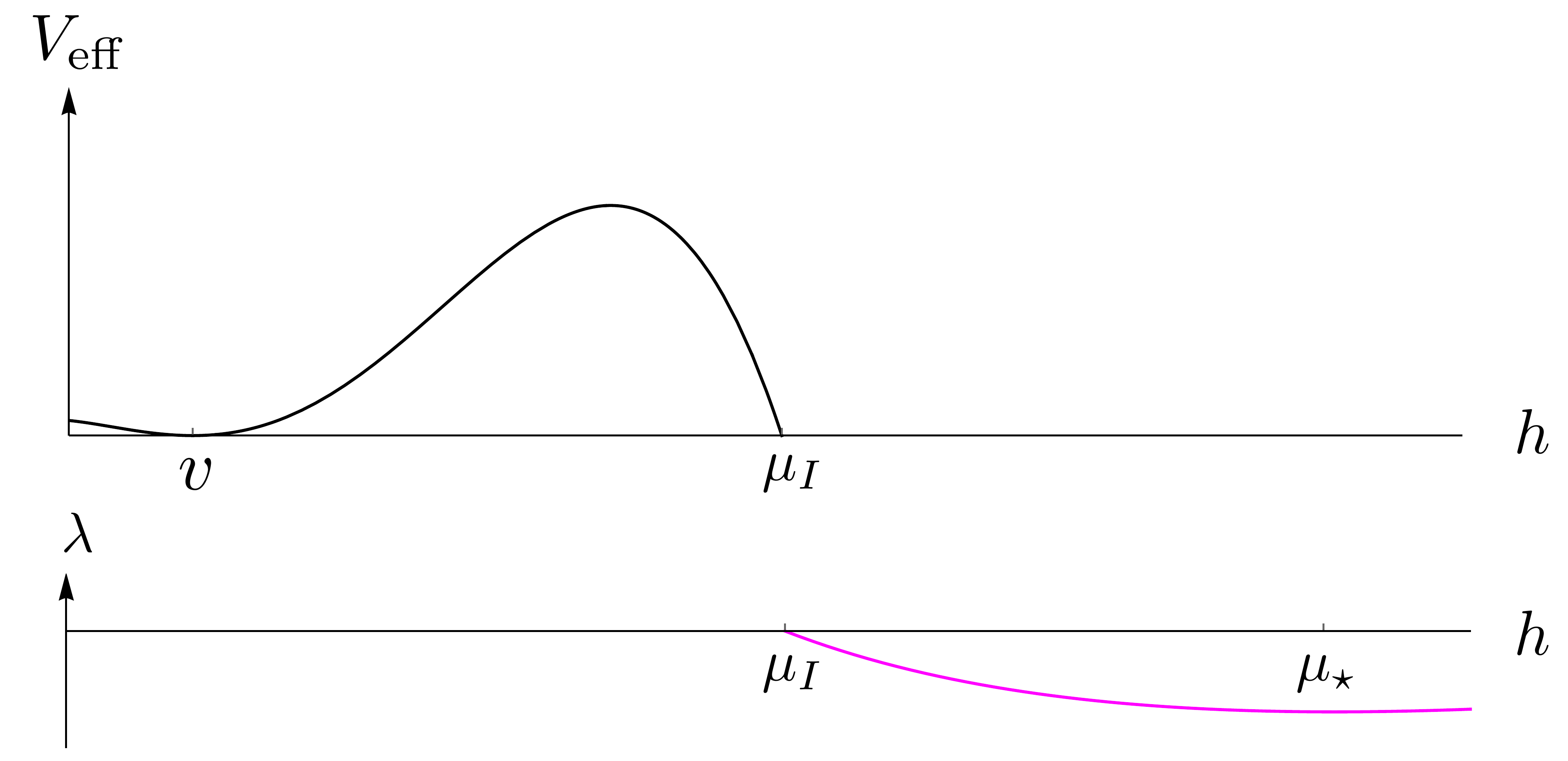}  
		\caption{The effective potential as well as the running of the quartic coupling, not to scale. If the scale~$\mu_\star$ where the quartic coupling reaches a minimum lies significantly below the Planck scale, then~$\mu_S \simeq \mu_\star$.}
		\label{mass bound sketch}
	\end{figure}
	Unfortunately,~\eqref{minst} on its own is insufficient to explain the large hierarchy between the electroweak scale and the Planck scale, for two reasons. First, it only shifts the question to why the instability scale lies so far below the Planck scale, and why it exists in the first place. Second, it only requires the Higgs mass to lie slightly below the instability scale, which as mentioned earlier is roughly~$10^{11}$ GeV, leaving unexplained an additional eight orders of magnitude to reach the electroweak scale. 
	
	The goal of this analysis is to address these two points as follows. With regards to the first point, as mentioned earlier, we assume the existence of the instability scale to be the result of some underlying UV physics determining the Higgs' couplings at some high energy. To achieve metastability, the quartic coupling must become negative at the instability scale~$\mu_I$, and then continue to fall off until it reaches a sufficiently negative value~$\lambda (\mu_S)$ to obtain the desired lifetime. Since~$\lambda$ depends only logarithmically on the scale, this running stretches over several orders of magnitude. This, together with relation~\eqref{minst}, provides us with an upper bound for the Higgs mass parameter and therefore also the electroweak scale, as sketched in Fig.~\ref{mass bound sketch}. As to the second point, in Sec.~\ref{massive nu's sec} we explore beyond-the-SM physics naturally lowering the instability scale, and thereby strengthening the bound on the Higgs mass. 
	
	We first derive a simple analytical relation between the vacuum's lifetime and the strength of the mass inequality, showing that less stable vacua lead to a stronger bound on the Higgs mass.
	
	For the purpose of an approximate analytical treatment, it suffices to use the leading-order expression for the decay rate:
	\begin{equation}
		\frac{\Gamma}{V} \propto \mu_S^4\, \exp \bigg(-  \frac{8 \pi^2}{3 |\lambda (\mu_S)|} \bigg)\,, 
		\label{ratesimple}
	\end{equation}
	where~$\mu_S$ satisfies~\eqref{saddle}. Note that, although the correction through either gravity or a dimension-6 term can be significant in minimizing~$S_{\rm E}$, the validity of the semi-classical description of course requires~$\mu_S \ll M_{\text{Pl}}$, or $\mu_S \ll \Lambda$ respectively, so that the direct correction to the numerical value of~$S_{\rm E}$ is small and can be safely neglected for the purpose of this estimate.
	
	Thus, at this level, the decay rate is fully determined by the two parameters~$\lambda (\mu_S)$ and~$\mu_S$. 
	For a given~$\mu_S$, it is easy to see that a larger~$\lambda (\mu_S)$ ({\it i.e.}, smaller~$|\lambda (\mu_S)|$) corresponds to a smaller decay rate, and hence a longer lifetime. Indeed, combining~\eqref{ltew} and~\eqref{ratesimple} gives 

	\begin{align}
		\nonumber
		\lambda (\mu_S) \simeq&  - \frac{8 \pi^2}{223.14+ 3 \ln \left(\frac{\mu_S^4}{H^3_0 \text{GeV}}\right) + 3 \ln \left(\frac{\tau_{\rm EW}}{\text{yrs}}\right)} \\ 
		\simeq &- \frac{8 \pi^2}{1551.15+12 \ln \left(\frac{\mu_S}{4 \times 10^{16}{\rm GeV}}\right)+ 3 \ln \left(\frac{\tau_{\rm EW}}{\text{yrs}}\right)}\,,
		\label{Invlifetime}
	\end{align}
	where in the last step we have substituted~$H_0 \simeq 10^{-42}$~GeV and normalized~$\mu_S$ to its SM value. Notice from the first line that, given the smallness of~$H_0$ compared to typical
	values of~$\mu_S$, the Hubble constant plays a significant role in determining the lifetime. This highlights the fact that metastability does not only require a conspiracy of SM couplings, but also of the Hubble and instanton scales. 
	
	The key point is that, once a lifetime~$\tau_{\rm EW}$ and an instanton scale~$\mu_S$ are specified, the quartic coupling~$\lambda (\mu_S)$ is determined through~\eqref{Invlifetime}. Further substitution
	of~$\mu_S$ and~$\lambda (\mu_S)$ back into~\eqref{saddle} then determines~$\beta_\lambda (\mu_S)$. This gives us all the necessary data to perform a RG evolution and determine the instability scale~$\mu_I$ at which~$\lambda$ crosses zero. While we will soon solve the RG equations numerically, it is worth plowing ahead with an analytical estimate to get some intuition on how different parameters of the theory ultimately affect the Higgs mass bound. 
	
	To proceed, let us Taylor-expand the quartic coupling around~$\mu_S$, and evaluate the result at~$\mu_I$:
	\begin{equation}
		\begin{split}
			0  = \lambda(\mu_I) =&   \lambda (\mu_S) + \beta_\lambda (\mu_S)\big(\ln \mu_I- \ln \mu_S\big) + \\
			&+ \frac{1}{2} \beta_\lambda^\prime (\mu_S)\big(\ln \mu_I- \ln \mu_S\big)^2 + \ldots
		\end{split}
		\label{Taylor}
	\end{equation}
	To leading order, this gives
	\begin{equation}
		\mu_I \simeq  \mu_S \exp \bigg(- \frac{|\lambda (\mu_S)|}{|\beta_\lambda (\mu_S)|}\bigg) \,.
		\label{LambdaImaxCH}
	\end{equation}
	In the regime where~$\mu_S \simeq \mu_\star$, which occurs whenever~$\mu_\star \ll M_{\text{Pl}}$ or ~$\mu_\star \ll \Lambda$ respectively, we must work to subleading order, since~$\beta_\lambda (\mu_\star) = 0$. The result in this case is an even larger hierarchy,
	\begin{equation}
		\mu_I \simeq \mu_\star \exp \bigg(- \sqrt{2 \frac{|\lambda (\mu_\star)|}{\beta_\lambda^\prime (\mu_\star)}} \bigg) \,. 
		\label{LambdaImax}
	\end{equation}
	For all parameter values of interest in our discussion, the right-hand sides of~\eqref{LambdaImaxCH} and~\eqref{LambdaImax} imply the existence of a significant hierarchy between the instability scale~$\mu_I$ and the instanton scale~$\mu_S$. This approximate result can now also be used to understand the dependence of our bound~\eqref{Ineq} on the Higgs mass on the lifetime. It is straightforward to see that, for a given value of~$\mu_S$, shorter lifetimes imply smaller values of~$\mu_I$, and thus stronger upper bounds on~$m_h$.

	\section{Metastability bounds in the Standard Model with gravity}
	
	\label{vac lifetime higgs mass numerics}
	
	As a first example for the consequences of our previous arguments, we assume that the SM is valid up to the Planck scale and neglect all Yukawa couplings except the top and bottom quark's and that of the tau. Furthermore, following~\cite{Buttazzo:2013uya} we treat~$y_{\rm t} (M_{\text{Pl}})$ and~$\lambda(M_{\text{Pl}})$ as free parameters characterizing the vacuum, while, for simplicity, fixing all other couplings at the Planck scale to their SM-extrapolated values. 
	
	We now perform the RG running numerically to accurately evaluate the Higgs mass bound~\eqref{Ineq}. This requires us to determine the instability scale~$\mu_I$ and all relevant couplings at that scale. 
	We assume that the SM is valid up to the Planck scale, and keep track of the most important coupling constants in determining the lifetime, namely the Higgs quartic, the top, bottom and tau Yukawa, and the three gauge couplings. We run these couplings using the complete three-loop beta functions for all but the small bottom and tau Yukawas and taking into account the most important four-loop contribution for the QCD coupling, given in the Appendix. 
	
	Following the reasoning of~\cite{Buttazzo:2013uya}, we assume~$y_{\rm t} (M_{\text{Pl}})$ and~$\lambda (M_{\text{Pl}})$ to be subject to the mechanism responsible for picking the vacuum and thus consider them as free parameters probing the landscape. While scanning different values of these parameters, we keep the gauge as well as the bottom and tau Yukawa couplings fixed at the Planck scale. We determine them by integrating the beta functions given Appendix~\ref{beta functions append} with the initial values of the relevant couplings at the top mass given in~\cite{Huang:2020hdv}, 
	\begin{widetext}
		
		\begin{gather}
			\lambda (M_t)=0.12607\,; \ \ y_t (M_t)=0.9312\,;   \ \ y_b (M_t)=0.0155334 \,;   \ \ y_\tau (M_t)= 0.0102566 \nonumber \\ 
			g_s (M_t)=1.1618\,; \ \   g^\prime (M_t)=0.358545\,; \ \    g(M_t)=0.64765 \,,
		\end{gather}
		leading to the Planck scale values
		\begin{gather}
			\lambda (M_{\text{Pl}})=-0.00971659\,; \ \ y_t( M_{\text{Pl}})=0.380196\,;   \ \ y_b (M_{\text{Pl}})=0.00456352 \,;   \ \ y_\tau (M_{\text{Pl}})= 0.00940597   \nonumber \\ 
			g_s (M_{\text{Pl}})=0.486966\,; \ \   g^\prime (M_{\text{Pl}})=0.477243\,; \ \   g(M_{\text{Pl}})=0.505385 \,.
		\end{gather}
		
	\end{widetext}
	Having solved for the RG evolution of these couplings, we determine~$\mu_S$ by solving~\eqref{saddle}, which in turn can be used to calculate the decay rate and thus the lifetime of the vacuum. For this purpose, we use the full expression~\eqref{ratefull} for the decay rate. Repeating this for a sufficiently tight mash of vacua, we obtain a collection of data points~$\left(y_{\rm t}(M_{\text{Pl}}),\lambda (M_{\text{Pl}}),\mu_S,\tau_{\rm EW},\overline{m}_h\right)$, where~$\overline{m}_h$ denotes the upper bound on the running Higgs mass for the considered vacuum. When scanning the landscape of potential vacua, we neglect those with a lifetime shorter than~$10^{10}$ years, as well as those whose dominant instanton reaches into a Planckian regime at its center,~$H_{R = \mu_S^{-1}} (r=0) \sim M_{\text{Pl}}$, for which quantum gravity effects might be relevant.
	
	We treat gravitational corrections in the way described in Sec.~\ref{EW vacuum decay}, taking into account only the leading order correction. This is justified, as the condition~$H_{R_S} \lesssim M_{\text{Pl}}$ ensures that the natural expansion parameter,~$\epsilon_{\text{grav}}= \frac{1}{\sqrt{\lambda (\mu_S)}}\frac{\mu_S^2}{M_{\text{Pl}}^2}$ is smaller than~$\frac{1}{8}$. Furthermore, as was argued in~\cite{Gravcorcor}, it can be expected that the higher-order gravitational corrections due to back-reactions should have a negligible influence on the decay rate. 
	
\begin{widetext}
		\begin{figure*}
			\centering
			\includegraphics[width=12.9cm]{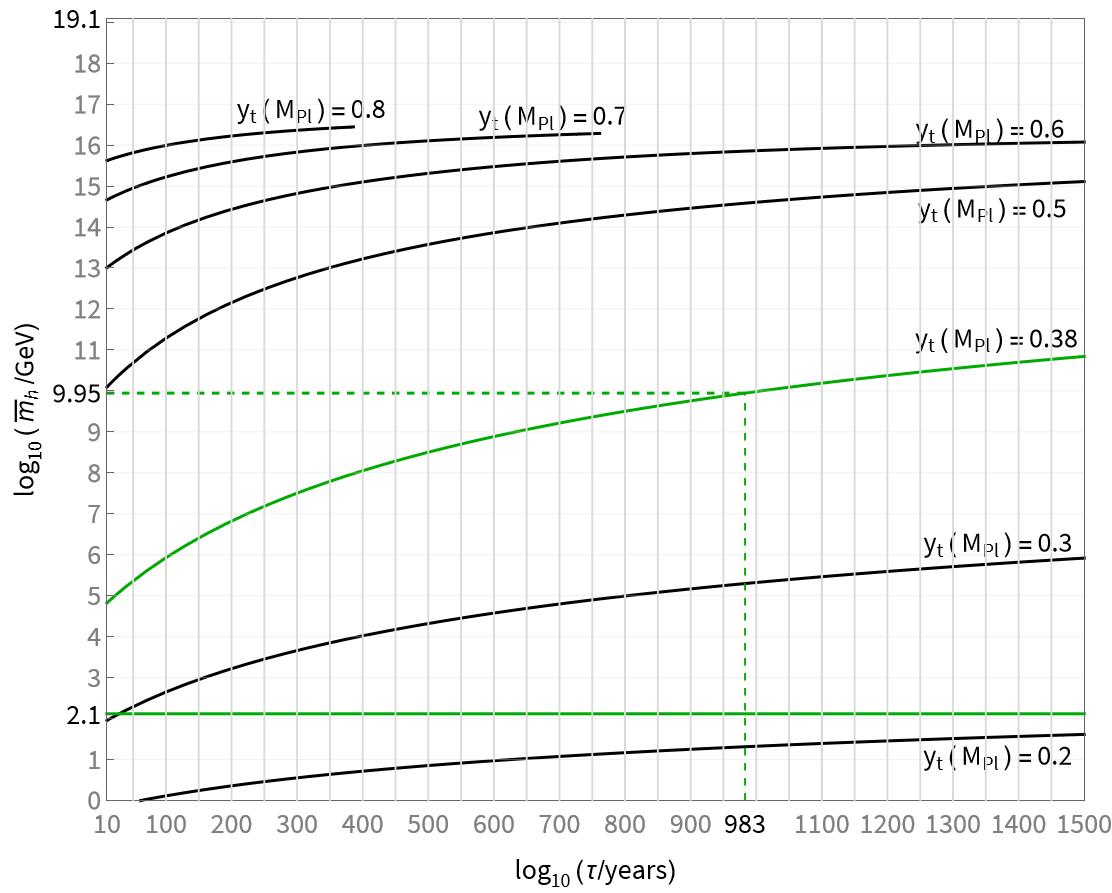}  
			\caption{The upper bound~$\overline{m}_h$ on the running mass as a function of the vacuum's lifetime for different values of~$y_{\rm t} (M_{\rm Pl})$. The dashed green line denotes the central values of parameters inferred from experiments, and the solid green line marks the measured value of the Higgs mass.} 	
			\label{mass bound SM plot}
		\end{figure*}
	\end{widetext}

	Figure~\eqref{mass bound SM plot} shows the numerical dependence of~$\overline{m}_h$ on the lifetime, where we replace~$\lambda (M_{\text{Pl}})$ by the lifetime and keep~$y_{\rm  t}(M_{\text{Pl}})$ as a second parameter. We find that once the lifetime has been imposed, a hierarchy of several orders of magnitude arises naturally, without the need for further fine-tuning of the couplings.
	Despite falling short of explaining the full hierarchy, this establishes a strong connection between metastability and the smallness of the Higgs mass. To summarize the logic, in any framework where there is a fundamental reason for vacuum metastability, such as in the early-time approach to eternal inflation, the requirement of an electroweak vacuum requires the running Higgs mass to be sufficiently small, with shorter lifetimes leading to stronger bounds.  
	
	As we will see throughout the remainder of this article, the insight that shorter lifetimes lead to a stronger bound on the Higgs mass remains valid even for the considered Standard Model extensions and is at least qualitatively independent of the choice of parameter(s) used to achieve different lifetimes. While this universality, which could be expected from the simplicity of~\eqref{LambdaImaxCH} and~\eqref{Ineq}, might be nothing but a mathematical peculiarity, we think that it is reasonable to believe that it might be of significance for the understanding of our universe, and in particular the Higgs mass.

	\section{Metastability bounds beyond the Standard Model}
	\label{massive nu's sec}

	The remaining gap between our bound and the observed value of the Higgs mass is mostly due to the hierarchy between the latter and the instability scale~$\mu_I$, which is not constrained by vacuum decay via an instanton. If, however, our vacuum's instability scale could be found at a lower energy, the bound arising from the requirement of metastability would improve significantly. In general, taking the idea seriously that the near-criticality of the Higgs mass is a consequence of the vacuum's metastability, this favors SM extensions causing large negative contributions to~$\beta_\lambda$ at energies significantly below the pure SM's instability scale. The simplest way to generate such terms is through the influence of relatively light fermions with a large Yukawa coupling to the Higgs. The currently best-motivated candidate for a particle with these properties is famously the right-handed neutrino\footnote{The case of a generic singlet neutrino has recently been discussed in \cite{Giudice:2021viw}, whereas we chose to focus on this particular example for concreteness, but also due to its potential phenomenological relevance.}. While we use is for the purpose of concreteness, it should be clear that, at least qualitatively, our results could be recovered from any suitable fermion.
	
	In generic see-saw models, the combination of relatively light right-handed neutrinos and a large Yukawa coupling is obviously in conflict with current experimental constraints, and the very idea of the see-saw itself. While it is in principle possible for the eigenvalues of the light neutrino mass matrix to be smaller than their pendants for the heavy neutrinos by many orders of magnitude, the cancellations necessary for this would require a significant amount of fine-tuning. An attractive way around this issue is presented by models in which the mass of the light neutrinos is protected by an approximate symmetry. For instance, the~$B-\Tilde{L}$ symmetric~$\nu$MSM~\cite{B-L,nufeat1,nufeat2,Asaka:2005an,nufeat3,nufeat4,nufeat5}, reviewed in subsection \ref{nu1}, allows for~$\mathcal{O}(1)$ couplings despite a right-handed neutrino mass as small as~$\mathcal{O}(1)$~TeV. 
	
	In subsections \ref{nu2}-\ref{nu4}, we demonstrate how adding light right-handed neutrinos to the SM decreases both the instability scale and the vacuum lifetime. However, as we show that the effect on the latter is far more pronounced than that on the instability scale, lowering it to the desired extent would render the vacuum severely unstable. 
	
	Again falling back to our initial assumption that the observed value of the Higgs mass is indeed the result of metastability, this suggests considering further SM extensions capable of conspiring with right-handed neutrinos in a similar way as gravity, {\it i.e.}, by stabilizing the vacuum at scales below the instability scale, thus allowing for stronger Yukawa couplings. The simplest way to achieve this is through new physics that generates a dimension-six term in the Higgs potential, which happens, {\it e.g.}, in models with a (partially) composite Higgs or two Higgs doublets like the MSSM. For a general dimension-six operator at tree-level, this mechanism and its interplay with the influence of fermion singlets has recently been discussed in \cite{Giudice:2021viw}. For concreteness and to allow for an efficient treatment of the running of the dimension-six Wilson coefficient, we will for our numerical analysis impose the matching condition of the SU(4)/Sp(4) composite Higgs model,
	\begin{equation}
		C_6 (\Lambda_f) = - \frac{\pi^2}{12} \lambda(\Lambda_f) \,,
		\label{C6 fix}
	\end{equation}
	where $\Lambda_f = 4 \pi f$, with the technipion decay constant $f$. It is however crucial to keep in mind that this case is generic, and our results can be easily generalized to a general theory.
	
	Using our results from section \ref{EW vacuum decay}, we show in subsections \ref{nu61} and \ref{nu62} that through the involvement of such an operator the instability scale can be lowered drastically, and with it the bound on the Higgs mass can be shifted closely above its observed value in agreement with all observational constraints. To be concrete, our perturbative treatment allows for a bound as strong as $10^4$~GeV in agreement with all observational constraints, and potentially even smaller if it weren't for the breakdown of approximation for the bounce.
	
	\subsection{Achieving light right-handed neutrinos with a large Yukawa coupling}\label{nu1}
	
	The $\nu$MSM consists of the SM together with three heavy, right-handed singlet neutrinos~$N_I$, $I = 1,2,3$. Their Lagrangian is given by
	\begin{equation}
		\mathcal{L}= \bar{N}_I i \gamma^\mu \partial_\mu N_I - Y_{\alpha I} \bar{\mathbf{L}}_\alpha N_I (\epsilon \mathbf{H}^*)- \frac{1}{2} M_{IJ} \bar{N}_I^c N_J + \text{h.c.} \,, 
		\label{LagrNeutr}
	\end{equation}
	where~$\bar{\mathbf{L}}_\alpha$ denotes the lepton doublets ($\alpha = e, \mu, \tau$),~$Y_{\alpha I}$ is the matrix of Yukawa couplings, and~$\epsilon$ is the totally anti-symmetric~SU(2) matrix. The last term is a Majorana mass matrix. This model is usually referred to as the~$\nu$MSM or type I see-saw, as the mass matrix of the left-handed neutrinos after symmetry breaking is given by
	\begin{equation}
		m_\nu = - \frac{v^2}{2} Y M^{-1} Y^T\,, 
		\label{neutrmass}
	\end{equation}
	which is to be understood as a matrix in generation space. This relation implies that, in generic cases, increasing the mass of the right-handed neutrinos decreases the mass of the left-handed ones, and vice versa. In other words, the lightness of the left-handed neutrinos is either the result of the large masses of their right-handed counterparts or tiny Yukawa couplings.   
	
	However, there is a subtlety to this argument. As the masses of the left-handed neutrinos are the eigenvalues of the mass matrix~\eqref{neutrmass}, it is possible that cancellations between the different elements of the latter lead to small eigenvalues despite Yukawa couplings of order~$\sim1$ and right-handed neutrinos with masses in the TeV-range~\cite{B-LMarko}. While the fine-tuning necessary to achieve such cancellations could of course be accidental, it can naturally arise as the result of an additional approximate symmetry protecting the masses of the left-handed neutrinos. One of the most important examples for such a symmetry is related to the preservation of the~$B-\Tilde{L}$ number, where~$\Tilde{L}$ denotes an extension of the lepton number by right-handed neutrinos~\cite{B-L}, and given by
	\begin{equation}
		N_3 \to e^{i \alpha} N_3, \ N_2 \to e^{- i \alpha} N_2, \ N_1 \to e^{i \beta} N_1 .
	\end{equation}
	Imposing this as an exact symmetry, and taking into account several observational constraints, the Majorana mass matrix and the matrix of Yukawa couplings are restricted to be of the form 
	\begin{equation}
		M =\begin{pmatrix}
			0 & 0 & 0\\
			0 & 0 & M\\
			0 & M & 0
		\end{pmatrix}\,;
		\qquad  
		Y= \begin{pmatrix}
			0 & Y_1 & 0\\
			0 & Y_2 & 0\\
			0 & Y_3 & 0
		\end{pmatrix}\,, \label{B-L}
	\end{equation}
	up to rotations of the right-handed neutrinos in flavor space. Consistency with observations requires this symmetry to be broken, which can be made manifest by introducing symmetry-breaking terms for both the Yukawa coupling as well as the mass term. While these are crucial to obtain everything from neutrino oscillations to the masses of the light mass eigenstates, {\it i.e.}, the SM neutrinos, their effect on the decay rate is negligible as long as the symmetry is only slightly broken. 
	
	The influence of right-handed neutrino singlets on vacuum stability has been discussed at different levels of detail in~\cite{nudecay&threshold,nudecay1,nudecay2,nudecay3,nudecay4}. Considering its strong dependence on the quartic coupling~$\lambda$, it should be clear that the neutrinos' main influence on the decay rate is via their contribution to~$\beta_\lambda$. While their precise effect depends on the detailed form of the Yukawa coupling and Majorana mass matrix, it is generally true that right-handed neutrinos lead to a more negative~$\lambda$, and thus a shorter lifetime.

	\subsection{Vacuum decay in symmetry protected see-saw models}\label{nu2}
	
	Assuming the Yukawa couplings to be of the form~\eqref{B-L} allows for the following replacements in the~$\nu$MSM's beta functions, which we give in full detail to two-loop accuracy in Appendix~\ref{beta functions append}:
	\begin{align}
		\nonumber
		Y_\nu Y_\nu^\dagger Y_\nu \rightarrow & Y_\nu \cdot \left(Y_1^2+Y_2^2+Y_3^2\right)\,; \\
		\text{Tr}\left(Y_\nu^\dagger Y_\nu\right)  \rightarrow & Y_1^2+Y_2^2+Y_3^2\,.
		\label{RHnu Yukawa replacements}
	\end{align}
	As argued above, observations require the addition of symmetry breaking terms to~\eqref{B-L}, and thus modifications to this relation. However, under our assumption that~$|Y| \sim \mathcal{O}(1)$ while~$M \lesssim 10^{11}$~GeV, such additional terms should be strongly suppressed compared to the leading-order terms, allowing us to safely neglect them in the following. The same is true for corrections of the heavy neutrino masses, so that we match the~$\nu$MSM with the pure SM at the scale~$\mu \simeq M$. This reduces the number of parameters to four: the heavy neutrino mass~$M$, and the three couplings~$Y_1(M),\ Y_2(M)$ and~$Y_3 (M)$.  
	
	Furthermore, given that we neglect the terms breaking the~$B-\tilde{L}$ symmetry, and taking into account that our analysis is in general insensitive to all properties of the neutrinos except their Yukawa couplings, we find an additional~$SO(3)$ symmetry among~$Y_1$,~$Y_2$ and~$Y_3$\footnote{Note that this symmetry would be broken, e.g., through the neutrino contribution to the running of the tau Yukawa. However, given the smallness of the latter as well as the smallness of the interval over which the neutrino contribution is relevant, we can safely neglect the corresponding terms for simplicity.}. This symmetry is also respected by the correction to the effective potential required for our upper bound on the Higgs mass, as well as by the threshold corrections at the matching scale of the pure SM's quartic coupling with the one of the UV theory, making it evident that the same holds for the NLO corrections to the decay rate from the neutrinos' determinants. Thus, as long as we are only interested in the neutrinos' influence on Higgs vacuum decay, we can further simplify the Yukawa matrix by rotating its components as 
	\begin{equation}
		Y\to  \begin{pmatrix}
			0 & 0 & 0\\
			0 & |Y| & 0\\
			0 & 0 & 0
		\end{pmatrix}\,, \ \text{where} \ |Y|^2= Y_1^2+Y_2^2+Y_3^2\,, 
		\label{YukMat}
	\end{equation}
	leaving us with just two parameters:~$M$ and~$|Y(M)|$\footnote{This is equivalent to the case of a hierarchy between the Yukawa couplings of the neutrinos, which is often used in the literature to simplify results~\cite{nudecay&threshold,nudecay1,nudecay2,nudecay3,nudecay4}.}. 
	
	The neutrino's contribution to the quartic coupling's loop correction,~$\lambda_1$, is given in, {\it e.g.},~\cite{nuSMmatching,nudecay3,nudecay4}, and the beta functions of the extended SM are known to two-loop accuracy~\cite{nuSMBeta2}. The only missing quantity to extend the NLO formula~\eqref{ratefull} is the neutrinos' kinetic term's functional determinant. Recalling that we are interested in the regime~$M \lesssim \mu_I$, the masses of both light and heavy neutrino states are negligible at scales relevant to the instanton. Making use of this fact, it is straightforward to see that the neutrino fluctuations contribute in the same way as the top quark under the replacement~$y_{\rm t}\rightarrow |Y|$.   
	
	\begin{widetext}
		\begin{figure*}
			\centering
			\includegraphics[width=12.9cm]{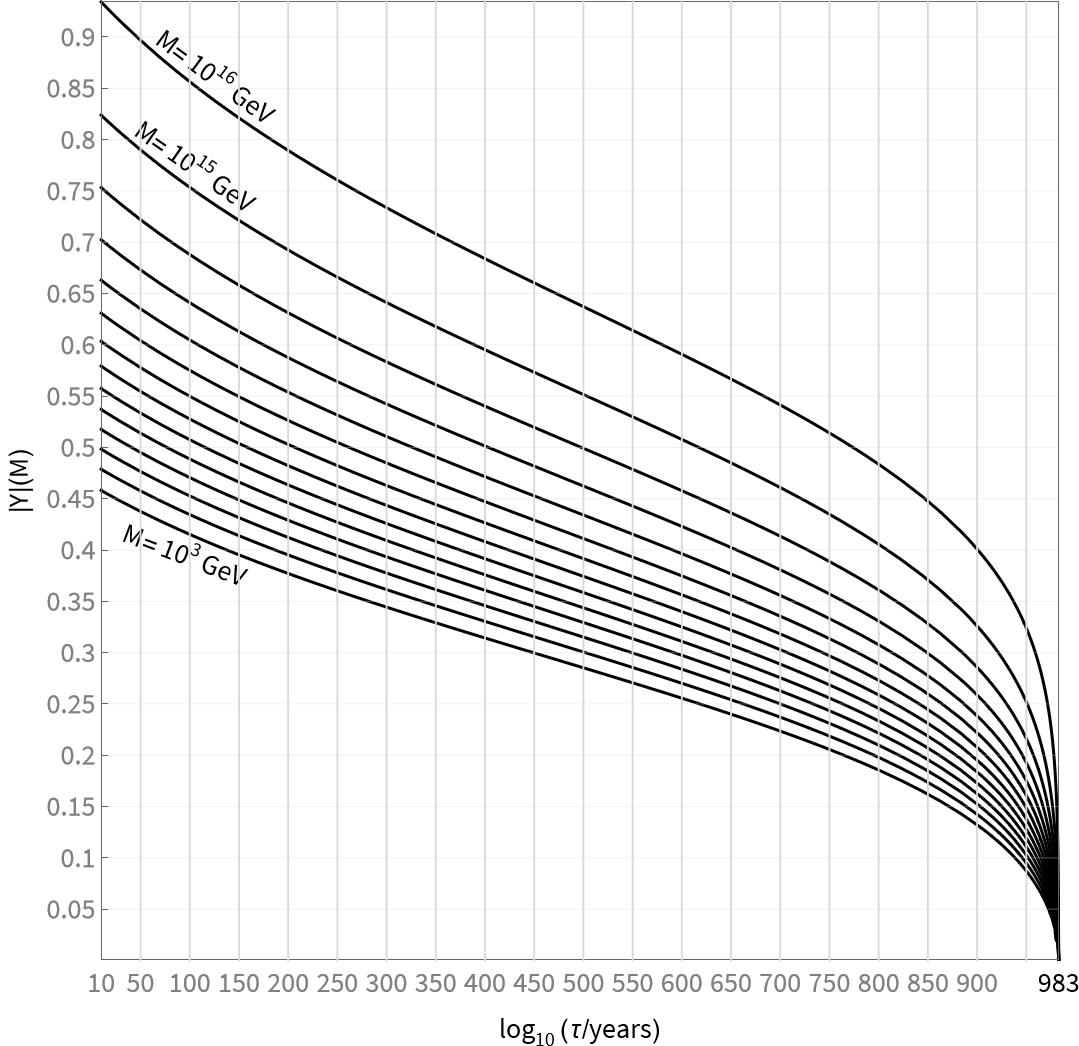} 
			\caption{The value of~$|Y(M)|$ necessary to realize a given lifetime, shorter than the central value of~$10^{983}$~years for the pure SM. Each curve corresponds to a different value of~$M$, in increasing order of magnitude.} 
			\label{rh neutrinos lifetime plot}
		\end{figure*}
	\end{widetext}
	
	\subsection{The numerical relation between neutrinos' Yukawa couplings and the lifetime}\label{nu3}
	
	We numerically compute the lifetime using~\eqref{ratefull}, supplemented with the neutrino determinant. To do so, we first match the SM couplings with those of the~$B-\tilde{L}$ symmetric~$\nu$MSM at the scale~$M$, following the Effective Field Theory procedures outlined in~\cite{nuEFT,nuEFT2} and using the threshold corrections for~$\lambda$ and~$y_{\rm t}$ given in~\cite{nudecay&threshold}\footnote{The authors of~\cite{nudecay&threshold} kindly informed us about a small typo in their Eq.~G(20), where a factor of~$\frac{1}{4}$ in the first term has to be replaced by~$\frac{1}{2}$. We have taken this into account in our calculation.}. Then, for a given value of~$|Y(M)|$, we run the couplings up to the Planck scale using the full~$\nu$MSM beta functions at two-loop accuracy found in~\cite{nuSMBeta2}, which we give in the Appendix in the conventions used in this article.  
	
	Having determined the running of the couplings, we obtain the instanton scale~$\mu_S$ as before by solving~\eqref{saddle}. It is worth noting that~$\mu_S$ lies significantly below~$\mu_\star$ (where~$\beta_\lambda$ vanishes) in this case. Indeed, the right-handed neutrino Yukawa couplings with values relevant for our purposes generally push~$\mu_\star$ beyond the Planck scale. Nevertheless, as explained in Sec.~\ref{EW vacuum decay}, gravitational corrections keep~$\mu_S$ below the Planck scale due to their non-logarithmic dependence on the scale.
	
	Figure~\ref{rh neutrinos lifetime plot} shows the value of~$|Y(M)|$ necessary to achieve a certain lifetime for all values of~$M$ of interest. We consider lifetimes ranging from~$10^{10}$~years to the SM central value of~$10^{983}$~years. This plot allows to convert any prediction of the vacuum's lifetime to a relation of the relevant right-handed neutrinos' properties, assuming either a strong hierarchy between the Yukawa couplings or the~$B-\tilde{L}$ symmetry. As a side result, we provide updated stability bounds on the Yukawa couplings, which can be read off from the left boundary of Fig.~\ref{rh neutrinos lifetime plot}. For~$M= 10^{12}-10^{15}$ GeV, our results are almost identical to the ones given in~\cite{nudecay&threshold}, with a minor deviation due to our more careful treatment of gravitational corrections and, more importantly, updated experimental input for the SM couplings. 
	\begin{widetext}
		\begin{figure*}
			\centering
			\includegraphics[width=12.9cm]{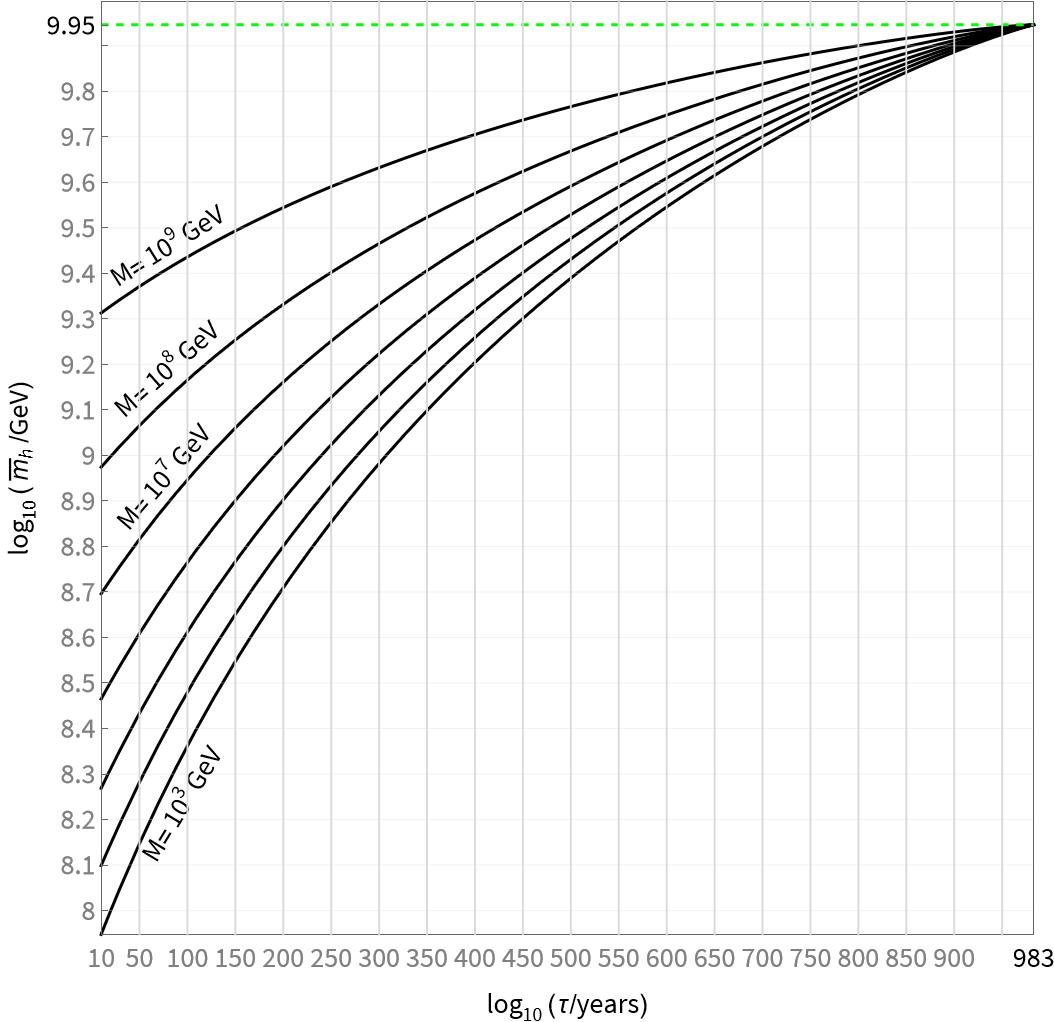}  
			\caption{The upper bound on the running Higgs mass as a function of the lifetime. Different lifetimes are achieved by varying the neutrino coupling parameter~$|Y(M)|$ (per Fig.~\ref{rh neutrinos lifetime plot}), while all other couplings are fixed to their observed values near the electroweak scale. Each curve corresponds to a different value of~$M$. The green dashed line marks the bound obtained for the considered set of parameters in the pure SM. }
			\label{rh neutrinos higgs bound plot}
		\end{figure*}
	\end{widetext}
	
	Before proceeding, it is important to stress once again that the inclusion of right-handed neutrino Yukawa couplings with properties relevant for our purpose generally shifts the minimum of~$\lambda$ to scales beyond the Planck mass. While this on its own would render the most important instantons inaccessible to us, we are saved by gravitational corrections. As described in Sec.~\ref{EW vacuum decay}, due to their non-logarithmic dependence on the scale, gravitational corrections shift~$\mu_S$ below the Planck scale, even though the gravitational term is strongly suppressed compared to the Minkowski contribution. This observation is crucial, as it provides a hint at how the small Higgs mass we observe might be achieved via a metastability bound.

	\subsection{The neutrino Yukawa couplings' effect on the metastability bound}\label{nu4}
	
	Having established the possibility to achieve any desired lifetime shorter than in the pure SM, we turn to our second objective, namely to address the smallness of the electroweak scale. Unlike in Sec.~\ref{higgs mass upper bound}, we keep~$\lambda$ and~$y_{\rm t}$ fixed to their SM values at the electroweak scale and instead scan the lifetime by varying~$|Y(M)|$. We further modify our previous procedure by including the neutrino correction to the decay rate, as well as the neutrino contributions to~$\lambda_1$ and the beta functions. We focus on the range~$M \lesssim \mu_I$, for otherwise the running of~$\lambda$ would be unaffected up to the instability scale and the Higgs mass bound would be unchanged.  
	
	Figure~\ref{rh neutrinos higgs bound plot} shows the upper bound on the running Higgs mass,~$\overline{m}_h$, as a function of the lifetime for different values of~$M$. Stronger Yukawa couplings lead to a faster decline of~$\lambda$, and thus a smaller~$\mu_I$, which in turn yields a stronger bound on the Higgs mass. Furthermore, it is clear that the larger the neutrino mass, the higher is the energy at which their Yukawa couplings become relevant, and the smaller is their influence on~$\mu_I$.  
	
	Our results lie within with experimental bounds on neutrino parameters. While in the $\nu$MSM many of their properties depend on the symmetry breaking parts of the Yukawa couplings and the mass matrix, the symmetric Lagrangian~\eqref{LagrNeutr} allows for a mixing of the active neutrinos with the Dirac fermion~$\Psi=N_2+N_3^c$, the strength of which is determined by the combination~$\frac{v}{M}|Y|$~\cite{B-LMarko}. Although experiments do not directly constrain this combination, current experimental bounds allow for~$|Y| \lesssim 0.5$ for~$M = 1$~TeV~\cite{NExp}, \textit{i.e.}, the full range of parameters investigated in this section.

	There is an important subtlety arising from the inclusion of neutrinos, or additional fermions in general. As we are mostly interested in scenarios with $M <\mu_I \sim \overline{m}_h$, the restricted Higgs mass is in general \textbf{not} the SM parameter, but that of the $\nu$MSM, which is related to its SM counterpart via a threshold correction $\delta m_h^2 = \frac{ M^2 |Y|^2}{(4 \pi)^2}$. While this expression is at least one order of magnitude smaller than $\overline{m}_h^2$ for all parameters considered in this article and thus negligible in the context our bound, it is important to note that this is not automatically true for all setups. The smallness of the correction compared to $\overline{m}_h^2$ can be related to the relative smallness of $|Y|$, which on the one hand leads to a small correction, and on the other hand does not allow for a fast enough decline of $\lambda$ to push $\mu_I$ close to $M$. This last point is further enhanced by the threshold correction in $\lambda$. If the matching scale $M$ lies close enough to the would-be instability scale, the threshold correction for $\lambda$ can be large enough for the latter to jump to a negative value without vanishing exactly, spoiling our analysis. While also this poses no problem for the range of parameters covered in this article, it might have consequences for a refined analysis of the model discussed in the next subsection with parameters currently not accessible by our perturbative treatment.

	\subsection{The numerical relation between neutrinos’ Yukawa couplings, dimension-six operator and the lifetime}\label{nu61}

	\begin{figure*}
		\centering
		\includegraphics[width=12.9cm]{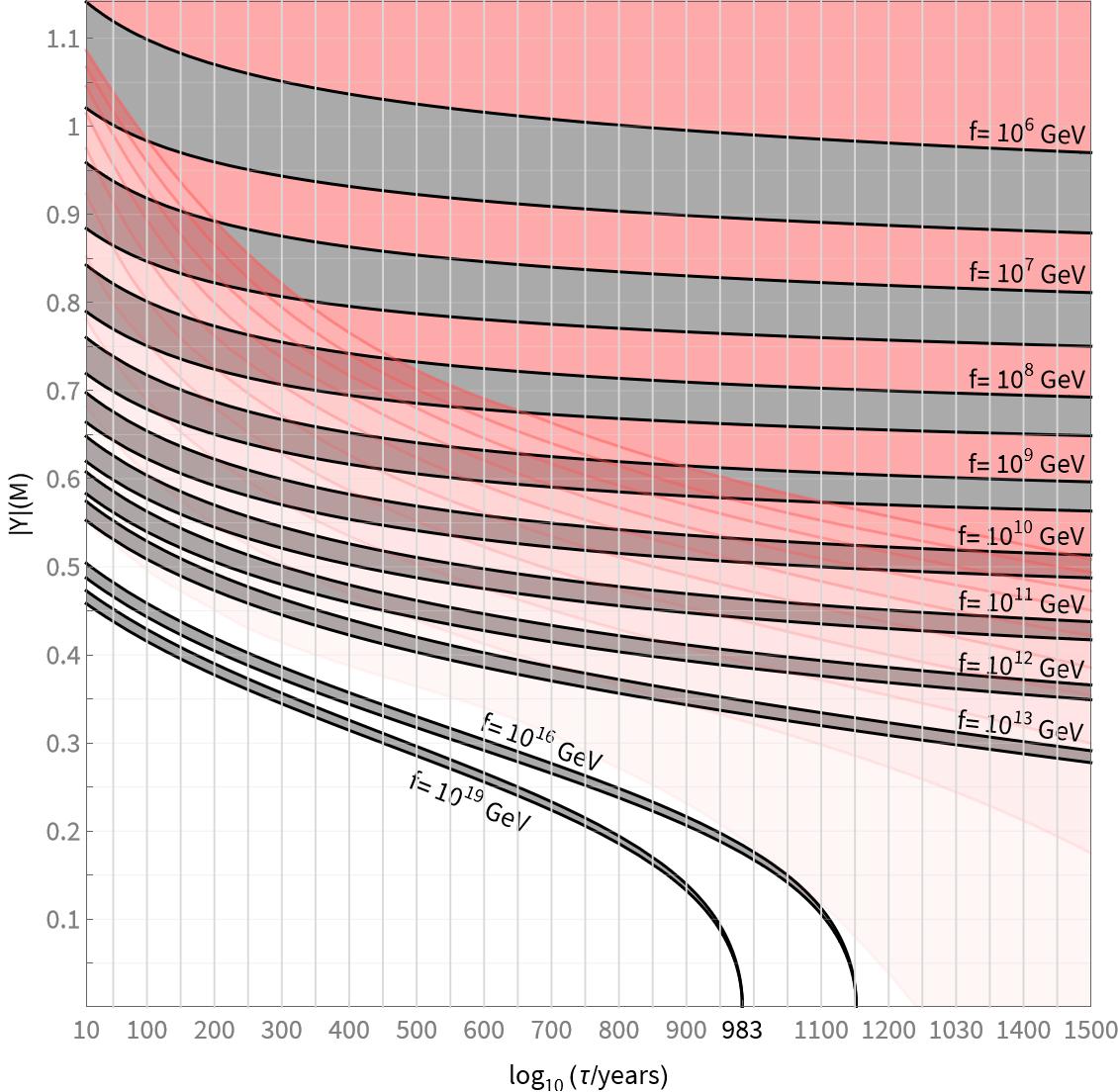}  
		\caption{The value of~$|Y(M)|$ necessary to realize a given lifetime for different values of~$f$. The gray bars cover the range~$M= 1-5$~TeV. In the white region, the expansion parameter defined in~\eqref{eps def} is smaller than~$0.1$ for $M=1$~TeV, in the dark red region larger than~$1$. The shaded regions indicate the transition in steps of~$\Delta \epsilon= 0.1$.}
		\label{lifetime CH + RHnu's}
	\end{figure*}

	Combining our previous results, it is straightforward to include the effect of a dimension-six operator. We restrict ourselves to small values of~$M$, for concreteness~$M \sim 1-5$~TeV, which yield the tightest bound on the Higgs mass (see Fig.~\ref{rh neutrinos higgs bound plot}) while being safely compatible with experimental constraints for~$M \gtrsim 2$~TeV. With these values fixed, we scan different values of~$f$ and~$|Y(M)|$. 
	
	All SM couplings are matched at the scale~$M$ using the same procedure as before, in particular taking into account threshold corrections for~$\lambda$ and~$y_{\rm t}$, and run using their three-loop beta functions, neglecting the contributions~$\sim m_h^2/\Lambda_f^2$. As in the most interesting cases the instanton scale lies close to the scale of new physics, we take into account the RG-running of $C_6$ at one-loop only.
	
	In this manner, we can compute the lifetime of the vacuum for each pair of values~$|Y(M)|$ and~$f$ --- or, equivalently, the value of~$|Y(M)|$ corresponding to every possible lifetime, given~$f$. This relation is depicted in Fig.~\ref{lifetime CH + RHnu's}, for values of~$f$ ranging from~$10^6$ GeV to $10^{19}$~GeV. In principle, smaller values of $f$ would also be possible, but are not accessible by our perturbative analysis. The red shading indicates different values of the expansion parameter~$\epsilon$ assuming $M=1$~TeV, defined in~\eqref{eps def}, ranging from~$0.1$ to~$1$. Consistent with the discussion below~\eqref{eps def}, we see that the perturbative expansion is indeed most reliable for shorter lifetimes and larger~$f$.

	\begin{figure*}
		\centering
		\includegraphics[width=12.9cm]{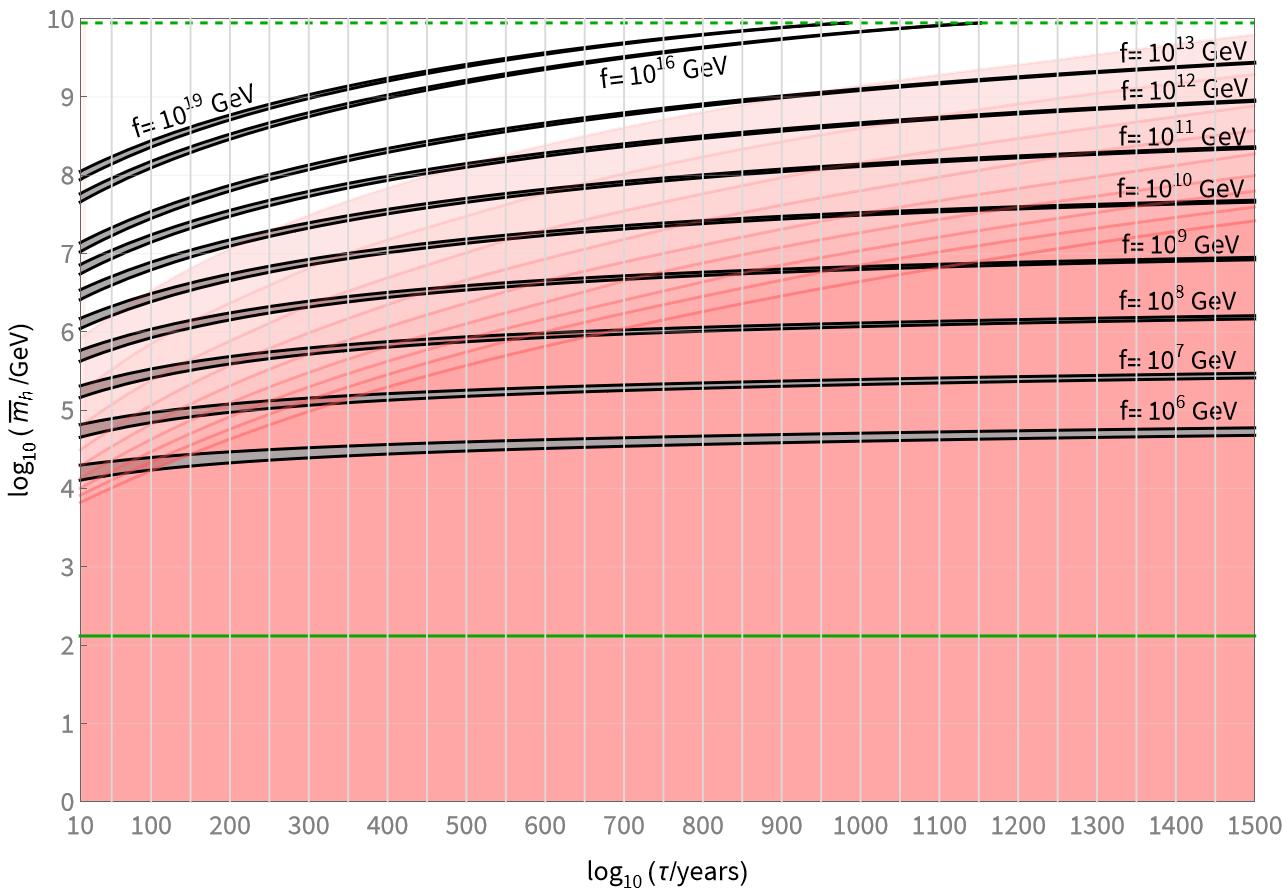}  
		\caption{The upper bound on the running Higgs mass as a function of the lifetime for different values of~$f$, with the range of $f$ being only restricted by the applicability of our perturbative treatment and not physical reasons. The gray bars represent the interval~$M=1-5$~TeV. The shading in the background again marks areas of different~$\epsilon$ for $M=1$~TeV, ranging from~$0.1$ to~$1$ in steps of size~$0.1$.}
		\label{mass bound CH + RHnu's}
	\end{figure*}

	\subsection{The neutrino Yukawa couplings’ and dimension-six term's combined effect on the metastability bound}\label{nu62}
	
	Having related the lifetime to the neutrino Yukawa couplings, we have gathered everything necessary to include the effect of the dimension-six term on the upper bound on the running Higgs mass as a function of the vacuum's lifetime. The modified relation is shown in Fig.~\ref{mass bound CH + RHnu's}, again considering~$M=1-5$~TeV for concreteness. We see that metastability forces~$m_h$ to lie consistently at least 2 orders of magnitude below~$f$, which is of the same order as the natural value for the mass parameter.
	
	Figure~\ref{mass bound CH + RHnu's} confirms our expectations. By balancing right-handed neutrinos with the stabilizing effect of a dimension-six operator, it is possible to simultaneously achieve a significantly smaller lifetime and decrease the upper bound on the running Higgs mass down to~$\simeq 10$~TeV. From there, naively extrapolating our results beyond the applicability of our approximations, it appears as if lowering $f$ just slightly further might allow for an upper bound on the Higgs mass of order 1~TeV, which, together with the necessity of a potential barrier separating electroweak and true vacuum, might be sufficient to explain the observed value of $10^2$~GeV as well as the remaining small hierarchy.

	\section{Conclusions}
	
	The absence of new physics in flavor, precision and LHC experiments points to the SM being valid up to very high energy, leaving the lightness of the Higgs unexplained and seemingly fine-tuned. It appears increasingly doubtful that the cherished principle of naturalness, which has guided much of particle model building, can explain the gauge hierarchy problem. A puzzling consequence of extrapolating the SM to very high energy scale is the metastability of the electroweak vacuum. We believe this numerical conspiracy is no accident. It is striking that Higgs metastability, the gauge hierarchy problem and the cosmological constant problem can all be interpreted as problems of near criticality.  
	
	In this article we considered an unconventional approach to the hierarchy problem, which rests on the idea that the gauge hierarchy is a consequence of a more primitive property of our vacuum, namely its metastability. 
	This is motivated, {\it e.g.}, by the early-time framework for eternal inflation put forth recently, based on search optimization on the string landscape. A key prediction of this framework is that optimal regions of the landscapes are characterized by vacua that are relatively short-lived, with lifetimes of order their de Sitter Page time. As such, this offers a {\it raison d'\^{e}tre} for the conspiracy underlying Higgs metastability. While we are primarily motivated by the non-equilibrium approach to eternal inflation, our results pertain more generally to any theoretical framework predicting that our vacuum should be metastable. 
	
	Central to our analysis is the observation that the metastability of the electroweak vacuum, together with the very requirement that such a non-trivial vacuum exists, implies that the Higgs mass is bounded from above by the instability scale. Simply put, a small Higgs mass is a necessary condition for a metastable vacuum. This was first pointed out in the context of the SM in~\cite{Buttazzo:2013uya}, though the bound is quite weak in this case, as the instability scale is~$\sim 10^{11}$~GeV.  
	
	A key point of this article is that simple, well-motivated extensions of the SM can significantly tighten the upper bound on the Higgs mass, as low as~$\sim 10$~TeV and potentially even further through a more elaborate analysis, by lowering the instability scale~$\mu_I$.  
	Furthermore, the vacuum lifetime can be shortened, to the extent that the Page time can be achieved. In other words, our viewpoint is that tightening the Higgs mass bound down to a value slightly above the electroweak scale, together with achieving a shorter lifetime, act as guiding principles in sifting through possible SM extensions. As an important byproduct of our analysis, we provided accurate calculations of the vacuum's lifetime as a function of the parameters characterizing the SM extensions of interest, which allowed us to update existing stability bounds by combining for the first time all relevant NLO corrections, both from gravity and functional determinants, at up to 3-loop accuracy.  
	
	We first considered the inclusion of right-handed neutrinos, a necessary extension to the SM to explain the mass of their left-handed counterparts. With a right-handed mass of order~TeV and~$\mathcal{O}(1)$ Yukawa couplings, the lifetime is dramatically shorter while the Higgs mass bound is significantly tighter. An elegant model where this can be naturally realized while satisfying experimental constraints is the~$\nu$MSM model with approximate~$B-\Tilde{L}$ symmetry. However, we found that right-handed neutrinos by themselves cannot fully explain the gauge hierarchy --- the tightest upper bound compatible with experimental constraints is~$\sim 10^8$~GeV --- still a few orders of magnitude away from the electroweak scale. Within the~$\nu$MSM, this discrepancy can only be alleviated by further increasing the Yukawa couplings, but this would inevitably render the electroweak vacuum unstable.   
	
	This observation led us to consider further SM extensions that have a stabilizing effect on the vacuum, thereby allowing for stronger right-handed neutrino Yukawa couplings. For concreteness, we studied the  minimal~SU(4)/Sp(4) composite Higgs model as a simple example that can achieve this. The main impact of compositeness is through the perturbative addition of a dimension-six correction to the Higgs potential. As such, while we focus on this particular minimal composite Higgs model, our results apply more generally to any theory that can be well-approximated by a dimension-six operator at low energy.  
	
	The stabilizing effect of this dimension-six operator allowed for stronger right-handed neutrino Yukawa couplings. This results in shorter lifetimes, of order the de Sitter Page time, while simultaneously lowering the upper bound on the Higgs mass. Within the reach of our perturbative analysis, the bound can be moved down to up to~$\simeq 10$~TeV, and potentially even further if corrections to the bounce profile were to be taken into account to increase our analysis' range of applicability.
	
	It is crucial to keep in mind that our result is really a bound  rather than an independent solution to the gauge hierarchy problem, and is independent of the mechanism setting the lifetime. The existence of our bound provides a strong constraint for any such mechanism, as it implies that its applicability either requires a small Higgs mass, or, ideally, should be able to yield the correct Higgs mass. Our result can, however, also be understood as a strong indicator for the shared origin of the different fine-tunings observed in the Higgs sector, as it suggests that \textbf{a} large hierarchy between running Higgs mass and Planck scale is necessary for metastability.
	
	The upper bound on the Higgs mass relies on two assumptions: i)~metastability of the electroweak vacuum; ii)~the existence of such a non-trivial vacuum, {\it i.e.}, a negative Higgs mass-squared. While the former is motivated, {\it e.g.}, by the non-equilibrium approach to eternal inflation, the latter remains an assumption in our analysis. In future work we will consider further SM extensions which naturally give rise to a negative Higgs mass-squared. The~$\nu$MSM model with approximate~$B-\Tilde{L}$ symmetry, with right-handed neutrino masses at the TeV scale, has the appealing feature of having the electroweak scale as its unique mass scale. A natural extension of the present work would be to consider classically conformally invariant models~\cite{Bardeen:1995kv,Hempfling:1996ht,Meissner:2006zh}, such as theories where scale invariance is radiatively broken in a hidden sector and mediated to the SM via a Higgs-portal coupling~\cite{Hempfling:1996ht,Chang:2007ki,Foot:2007as,Foot:2007iy,Iso:2009ss,Alexander-Nunneley:2010tyr,Iso:2012jn,Englert:2013gz,Chun:2013soa,Heikinheimo:2013fta,Hambye:2013dgv,Khoze:2013oga,Carone:2013wla,Khoze:2013uia,Steele:2013fka,Wang:2015cda}. It will be interesting to see whether such extensions can further tighten the bound closer to the measured Higgs mass.

	\section*{Acknowledgments}
	We thank Dario Buttazzo, Oscar Cata, Marco Drewes, Gian Giudice, Matthew McCullough, Andrey Pickelner, Francesco Sannino, Tevong You and Matthew Schwartz for helpful correspondence, and Jan-Niklas Toelstede and Sam Wong for many enlightening discussions. We furthermore thank the anonymous referee of Physical Review D for invaluable input, which helped us significantly improve this article. T.S. is funded by a PhD scholarship of the German federal ministry of science and education, provided by Heinrich-Böll foundation, which he thanks for years of great support. J.K. is supported by the US Department of Energy (HEP) Award DE-SC0013528, NASA ATP grant 80NSSC18K0694, and by the Simons Foundation Origins of the Universe Initiative.

	\appendix
	
	\section{Appendix}
	
	\subsection{Upper bound on the Higgs mass}
	\label{upper bound m}
	
	In this Appendix we derive the more precise version of the Higgs mass bound, given by~\eqref{Ineq}, by including non-logarithmic one-loop corrections to~$\lambda$. 
	The full RG-improved effective potential is given by 
	\begin{equation}
		\begin{split}
			V_{\text{eff}}(H)=& - \frac{m^2_h}{4} {\rm e}^{2 \Gamma[H]} H^2 + \\
			&+\frac{1}{4} \Big( \lambda (H) + \lambda_1 (H) + \ldots\Big) {\rm e}^{4 \Gamma[H]} H^4\,,
			\label{Vefffull}
		\end{split}
	\end{equation}
	where~$\lambda_n$ denotes non-logarithmic corrections to~$\lambda$ at~$n$-loop order. The extrema of this potential, $\left.\frac{\rm d}{{\rm d}H}V_{\text{eff}}(H)\right\vert_{H=v}=0$, 
	are given by
	\begin{equation}
		\begin{split}
			\frac{m^2_h}{v^2}= \frac{{\rm e}^{2 \Gamma (v)}}{1+\gamma (v)} \bigg(&2 \big(\lambda (v) +\lambda_1 (v) + \dots\big)+ \\
			&+\frac{1}{2}\big(\beta_\lambda (v) + \beta_{\lambda_1} (v) + \dots\big) \bigg)\, ,
			\label{Mineq}
		\end{split}
	\end{equation}
	where~$\gamma$ denotes the Higgs field's anomalous dimension.
	
	The existence of a metastable vacuum is equivalent to the existence of solutions to this equation. A first restriction on these solutions can be read off directly from~\eqref{Vefffull}: At scales somewhat larger than the instability scale,~$\lambda (\mu)$ becomes sufficiently negative to compensate for the positive loop corrections~$\{ \lambda_n (\mu) \}_n$. Thus at those scales both the quadratic and the quartic term appear with a negative sign, and no extremum can occur. This implies that we can focus on values of~$v$ near the instability scale, with the immediate consequence that~$\lambda (v) \simeq \beta_\lambda (\mu_I) \ln \frac{v}{\mu_I}$. Thus~$\lambda$ becomes effectively one-loop~\cite{Consistent}, and therefore a consistent leading-order perturbative expansion of~\eqref{Mineq} must into account all one-loop terms. This includes in particular~$\lambda_1$, which has been neglected in~\cite{Buttazzo:2013uya}. 
	
	Restricting ourselves to one-loop accuracy,~\eqref{Mineq} can be brought to the form
	\begin{equation}
		m^2_h= v^2 \bigg(2 \ln \frac{v}{\mu_I} + 2\frac{\lambda_1 (\mu_I)}{\beta_\lambda (\mu_I)}+ \frac{1}{2} \bigg) \beta_\lambda (\mu_I) \,.
		\label{m^2= append}
	\end{equation}
	Understood as a function of~$v$, the right-hand side is bounded from above since~$\beta_\lambda (\mu_I) <0$. Maximizing over~$v$, we obtain the inequality
	\begin{equation}
		m_h^2 \lesssim  \left\vert\beta_\lambda (\mu_I)\right\vert \exp\left(-\frac{3}{2} -2 \frac{\lambda_1 (\mu_I)}{\beta_\lambda (\mu_I)}\right)\mu_I^2\,.
		\label{CorrectBound}
	\end{equation}
	If~$\lambda_1$ is neglected, as assumed in~\cite{Buttazzo:2013uya}, this bound reduces to~\eqref{minst}. 
	
	Equation~\eqref{m^2= append} can furthermore be used to classify the effective potential. Clearly, for values of~$m_h^2$ larger than the bound~\eqref{CorrectBound}, it has no extremum. In the case that~$m_h^2$ is strictly smaller than the bound but larger than~$0$, the right hand side of equation~\eqref{m^2= append} has two solutions, corresponding to the top of the potential barrier and the vev, which --- neglecting the non-logarithmic term~$\lambda_1$ for now --- ranges from~$0$ to~${\rm e}^{- 1/4} \mu_I$. For~$v={\rm e}^{-3/4} \mu_I$ the inequality~\eqref{CorrectBound} is saturated, corresponding to a saddle point of the effective potential at~$v$, implying that corrections due to the running of~$\lambda$ effectively cancel the tree-level mass parameter~$m_h^2$ in~$\frac{\text{d}^2}{\text{d}H^2}V_{\text{eff}}(H)$. This is now precisely the critical point at the center of the probability distribution derived in \cite{Giudice:2021viw}, confirming its consistency with our bound on the running Higgs mass, which is saturates.
	
	For~$m_h^2<0$, the potential no longer permits spontaneous symmetry breaking without the influence of further quantum corrections and~\eqref{m^2= append} has only one solution, again corresponding to the potential wall, which is now located beyond the instability scale. Since in this limit the potential's only minimum lies at~$H=0$, we find~$v < {\rm e}^{- \frac{1}{4}} \mu_I$ when neglecting~$\lambda_1$. The same discussion can be performed after restoring it, leading to the more appropriate bound 
	\begin{equation}
		v\leq \exp\left(-\frac{1}{4} - \frac{\lambda_1 (\mu_I)}{\beta_\lambda (\mu_I)}\right)\mu_I
		\label{CorrectBoundVev}
	\end{equation}
	In Sec.~\ref{nu61} and \ref{nu62}, we add a dimension-six operator to the potential, of the form~$\Delta V= \frac{C_6}{\Lambda_f^2} H^6$, where to leading order~$C_6 \sim \lambda$. This term leads to a correction in~\eqref{CorrectBound} of order~$\lambda \frac{v^2}{\Lambda_f^2}$. Recalling that ~$v^2 \lesssim \mu_I^2 \ll \Lambda^2$, this correction is suppressed compared to the terms originating from the pure SM, so that it could justifiably be neglected in our perturbative discussion.

	\subsection{NLO tunneling formula for the Standard Model}
	For completeness, we include here the NLO formula for the vacuum decay rate derived in~\cite{AFS}. Before performing the integral over the dilatation modes, the decay rate per unit volume is given by
	\begin{equation}
		\frac{\Gamma}{V}=  \int \frac{\text{d}R}{R^5} \ {\rm e}^{- S_{\text{E}} \left(\lambda \left(R^{-1}\right),R\right)} D \big(R^{-1}\big)\,,
	\end{equation}
	The factor~$D \big(R^{-1}\big)$ is defined as
	\begin{widetext}
		\begin{align}
			\nonumber
			D \big(R^{-1}\big)  \equiv & \frac{72}{\sqrt{6}\pi^2} S_{\text{E}}^4 \left(\lambda \left(R^{-1}\right),R\right) \exp\left[12 \zeta^\prime (-1) - \frac{25}{3} + \pi^2 - \gamma_{\rm E} - \frac{3}{2} \ln 2 - \frac{3}{2} S_{\text{fin}}^+ \left(X\right) - 3 S_{\text{fin}}^+ \left(Y\right)  +\frac{3}{2} S_{\text{fin}}^{\bar{\psi}\psi} \left(\sqrt{Z_t} \right)  \right. \\
			& ~~~~~\left.   +\frac{3}{2} S_{\text{fin}}^{\bar{\psi}\psi} \left(\sqrt{Z_b} \right)-3\,S_{\text{loops}}^{\bar{\psi}\psi} \left(Z_t\right)-3\,S_{\text{loops}}^{\bar{\psi}\psi} \left(Z_b\right) -\frac{1}{2}S_{\text{diff}}^{\text{AG}}\left(X \right) - S_{\text{diff}}^{\text{AG}}\left(Y \right) - S_{\text{loops}}^{\text{AG}}\left(X \right) - 2 S_{\text{loops}}^{\text{AG}}\left(Y \right) \right] \,,
			\label{fullformula}
		\end{align}
		where~$X\equiv - \frac{g^2+{g^{\prime}}^2}{12 \lambda}$,~$Y \equiv -\frac{g^2}{12 \lambda}$, and~$Z_i \equiv \frac{y_{\rm i}^2}{\lambda}$. 
		The correction~$S_{\text{fin}}^+ (x)$ appearing in the exponent is given by
		\begin{align}
			\nonumber
			S_{\text{fin}} (x) =& x^2 \left(6 \gamma_{\rm E} +51-6\pi^2\right) + 6 x + \frac{11}{36} + \ln 2 \pi + \frac{3}{4 \pi^2}\zeta (3) - 4 \zeta^\prime (-1)  - \ln \left( \frac{\cos \left(\frac{\pi}{2} \kappa_x\right)}{6 \pi x} \right) \\
			\nonumber
			& -  x \kappa_x \left[ \psi^{(-1)} \left( \frac{3+\kappa_x}{2}\right)  - \psi^{(-1)} \left( \frac{3-\kappa_x}{2}   \right) \right] + \left( 6x - \frac{1}{6}\right) \left[ \psi^{(-2)} \left( \frac{3+\kappa_x}{2}\right) + \psi^{(-2)} \left( \frac{3-\kappa_x}{2}   \right) \right] \\ 
			& +  \kappa_x \left[ \psi^{(-3)} \left( \frac{3+\kappa_x}{2}\right) - \psi^{(-3)} \left( \frac{3-\kappa_x}{2}\right) \right]  - 2\left[ \psi^{(-4)} \left( \frac{3+\kappa_x}{2}\right) + \psi^{(-4)} \left( \frac{3-\kappa_x}{2}\right) \right]\,,
		\end{align}
		where~$\kappa_x \equiv \sqrt{1-24 x}$ and~$\psi^{n}$ is the polygamma function. The other corrections to the action are 
		\begin{align}
			\nonumber
			S_{\text{diff}}^{\text{AG}} (x) =& x^2 \left(121 - 12 \pi^2 \right) - \frac{45}{2} x^2\,; \\
			\nonumber
			S_{\text{loops}}^{\text{AG}} (x) =& - \frac{5}{18}  - \frac{1}{3} \left(\gamma_{\rm E}  - \ln 2\right) -   x \Big(7 + 6 (\gamma_{\rm E} - \ln 2) \Big) -   9 x^2 \bigg( \frac{1}{2} + \gamma_{\rm E} - \ln 2 \bigg)\,; \\ 
			S_{\text{loops}}^{\bar{\psi}\psi} (x) =& - x \bigg( \frac{13}{8} + \frac{2}{3} (\gamma_{\rm E} -  \ln 2) \bigg) + x^2 \bigg( \frac{5}{18} + \frac{1}{3}(\gamma_{\rm E} - \ln 2) \bigg) \,,
		\end{align}
		as well as 
		\begin{align}
			\nonumber
			S_{\text{fin}}^{\bar{\psi}\psi} (x) = & 16 \psi^{(-1)} (2) - \frac{8}{3} \psi^{(-2)} (2) + \frac{4}{3} x^2 (1-\gamma_{\rm E}) - \frac{x^4}{3} (1- 2 \gamma_{\rm E}) \\
			\nonumber
			&- \frac{4}{3}x (1-x^2) \bigg[ \psi^{(-1)}(2+x)- \psi^{(-1)}(2-x) \bigg] + \frac{4}{3}x (1-3 x^2) \bigg[ \psi^{(-2)}(2+x) +  \psi^{(-2)}(2-x) \bigg] \\ 
			&+8x \bigg[ \psi^{(-3)}(2+x)- \psi^{(-3)}(2-x) \bigg]-8 \bigg[ \psi^{(-4)}(2+x) +  \psi^{(-4)}(2-x) \bigg] \,.
		\end{align}
		We refer the reader to~\cite{AFS} for further details and the meaning of the different subscripts. \newpage
		
		\subsection{Beta functions}
		\label{beta functions append}
		
		We provide the beta functions used in our calculations. The beta function of the quartic coupling~$\lambda$ at 3-loop order is given by
		\begin{align}
			\nonumber
			\beta_\lambda =& \frac{1}{(4 \pi)^2}\left[ 24 \lambda^2 - 6 y_{\rm t}^4 - 6 y_{\rm b}^4  - 2 y_{\tau}^4 + \frac{3}{8} \left(2 g^4 + \left(g^2+{g^{\prime}}^2\right)^2 \right) - \lambda\left(9 g^2+ 3 {g^{\prime}}^2 - 12 y_{\rm t}^2 - 12 y_{\rm b}^2 - 4 y_{\tau}^2\right) +|Y|^2\left( 4 \lambda  - 2 |Y|^2 \right) \right]  \\
			\nonumber
			&+\frac{1}{(4 \pi)^4}\bigg[ \frac{1}{48} \left(915 g^6 - 289 g^4 {g^{\prime}}^2 -559 g^2 {g^{\prime}}^4 -379 {g^{\prime}}^6 \right) +30 y_{\rm t}^6 +30 y_{\rm b}^6 +10 y_{\tau}^6 -y_{\rm t}^4 \left( \frac{8}{3}{g^{\prime}}^2 +32 g_s^2 +3 \lambda +6 y_{\rm b}^2 \right)- \\ \nonumber
			&-y_{\rm b}^4 \left( \frac{4}{3}{g^{\prime}}^2 +32 g_s^2 +3 \lambda +6 y_{\rm t}^2\right)-y_{\tau}^4 \left( 4{g^{\prime}}^2 +3 \lambda \right)\\
			\nonumber
			&  ~~~~~~~~ + \, \lambda\left(- \frac{73}{8} g^4 +\frac{39}{4} g^2 {g^{\prime}}^2 + \frac{629}{24} {g^{\prime}}^4 +108 g^2 \lambda +36 {g^{\prime}}^2 \lambda -312 \lambda^2 \right)\\
			\nonumber
			& ~~~~~~~~  +\, y_{\rm t}^2 \left(- \frac{9}{4}g^4 + \frac{21}{2} g^2 {g^{\prime}}^2 - \frac{19}{4}{g^{\prime}}^4 + \lambda \left(\frac{45}{2} g^2 +\frac{85}{6} {g^{\prime}}^2 +80 g_s^2 -144 \lambda - 42 y_{\rm b}^2 \right)\right)\\
			\nonumber
			& ~~~~~~~~  +\, y_{\rm b}^2 \left(- \frac{9}{4}g^4 + \frac{9}{2} g^2 {g^{\prime}}^2 - \frac{5}{4}{g^{\prime}}^4 + \lambda \left(\frac{45}{2} g^2 +\frac{25}{6} {g^{\prime}}^2 +80 g_s^2 -144 \lambda - 42 y_{\rm t}^2 \right)\right)\\
			\nonumber
			& ~~~~~~~~  +\, y_{\tau}^2 \left(- \frac{3}{4}g^4 + \frac{11}{2} g^2 {g^{\prime}}^2 - \frac{25}{4}{g^{\prime}}^4 + \lambda \left(\frac{15}{2} g^2 +\frac{75}{6} {g^{\prime}}^2  -48 \lambda \right)\right)
			\\
			\nonumber
			& ~~~~~~~~  +\, |Y|^2\bigg( -96 \lambda^2  + \lambda (5 {g^\prime}^2  + 15 g^2  - 2 |Y|^2) -  \frac{3}{2} g^4 + 20 |Y|^4\bigg) \bigg]  \\ 
			\nonumber
			&+\frac{1}{(4 \pi)^6}\bigg[\lambda^3 \left(12022.7 \lambda + 1746 y_{\rm t}^2 - 774.904 g^2 - 258.3 {g^\prime}^2 \right) \\
			\nonumber
			&  ~~~~~~~~ +\, \lambda y_{\rm t}^2 \left(3536.52 y_{\rm t}^2 + 321.54 g_s^2 - 719.078 g^2 - 212.896 {g^\prime}^2\right) \\ 
			\nonumber
			&  ~~~~~~~~ +\, \lambda^2 \left(-1580.56 g^4 - 1030.734 {g^\prime}^4 - 1055.466 g^2 {g^\prime}^2\right) \\ 
			\nonumber
			&  ~~~~~~~~ +\,  \lambda y_{\rm t}^4 \left(-446.764 y_{\rm t}^2 - 1325.732 g_s^2 - 10.94 g^2 - 70.05 {g^\prime}^2\right)  \\ 
			\nonumber
			&  ~~~~~~~~ +\,  \lambda y_{\rm t}^2 \left(713.936 g_s^4 - 639.328 g^4 - 415.888 {g^\prime}^4 + 30.288 g_s^2 g^2 + 58.18 g_s^2 {g^\prime}^2 + 18.716 g^2 {g^\prime}^2\right) \\
			\nonumber
			&  ~~~~~~~~ +\,  \lambda g^4 \left(-114.288 g_s^2 + 1730.966 g^2 + 265.46 {g^\prime}^2\right) + \lambda {g^\prime}^4 \left(-46.562 g_s^2 + 343.072 g^2 + 260.814 {g^\prime}^2\right) \\ 
			\nonumber
			&  ~~~~~~~~ +\, y_{\rm t}^6 \left(-486.298 y_{\rm t}^2 + 500.988 g_s^2 + 146.276 g^2 + 113.1 {g^\prime}^2\right)  \\
			\nonumber
			&  ~~~~~~~~ +\, y_{\rm t}^4 \left(-100.402 g_s^4 + 31.768 g^4 + 88.6 {g^\prime}^4 + 26.698 g_s^2 g^2 + 58.566 g_s^2 {g^\prime}^2 - 234.52 g^2 {g^\prime}^2\right)  \\ 
			\nonumber
			&  ~~~~~~~~ +\,  y_{\rm t}^2 g_s^2 \left(32.928 g^4 + 3.644 {g^\prime}^4 + 37.954 g^2 {g^\prime}^2\right) + y_{\rm t}^2 g^4 \left(125 g^2 + 43.470 {g^\prime}^2\right)  \\ 
			\nonumber
			& ~~~~~~~~ +\,  y_{\rm t}^2 {g^\prime}^4 \left(58.318 g^2 + 102.936 {g^\prime}^2\right) + g_s^2 \left(15.072 g^6 + 7.138 {g^\prime}^6 + 5.024 g^4 {g^\prime}^2 + 6.138 g^2 {g^\prime}^4\right) \\
			& ~~~~~~~~  -\, 228.182 g^8 -    23.272 {g^\prime}^8 -  126.296 g^6 {g^\prime}^2 + 36.112 g^4 {g^\prime}^4 - 14.288 g^2 {g^\prime}^6 \bigg]  +  \mathcal{O} \left( \frac{m^2_h}{\Lambda^2_f} \right)  \,.
		\end{align}
		\newpage
		For the top Yukawa coupling~$y_{\rm t}$, we have, also at 3-loop order,
		\begin{align}
			\nonumber
			\beta_{y_{\rm t}} =& \frac{y_{\rm t}}{(4 \pi)^2}\bigg[ - \frac{9}{4} g^2 - \frac{17}{12}{g^{\prime}}^2 -8 g_s^2 + \frac{9}{2} y_{\rm t}^2 +|Y|^2 \bigg] \\ 
			\nonumber
			&+ \frac{y_{\rm t}}{(4 \pi)^4}\bigg[ - \frac{23}{4} g^4 - \frac{3}{4}g^2 {g^{\prime}}^2 + \frac{1187}{216} {g^{\prime}}^4 + 9 g^2 g_s^2 + \frac{19}{9} {g^{\prime}}^2 g_s^2 -108 g_s^4 +  \\
			\nonumber
			&  ~~~~~~~~ + y_{\rm t}^2 \left( \frac{225}{16} g^2 +\frac{131}{16} {g^{\prime}}^2 +36 g_s^2 - \frac{11}{4} y_{\rm b}^2 -\frac{9}{4} y_{\tau}^2 \right)\\ 
			\nonumber
			&  ~~~~~~~~ + y_{\rm b}^2 \left( \frac{99}{16} g^2 +\frac{7}{48} {g^{\prime}}^2 +4 g_s^2 -\frac{1}{4} y_{\rm b}^2 + \frac{5}{4} y_{\tau}^2\right) + y_{\tau}^2 \left( \frac{15}{8} g^2 +\frac{25}{8} {g^{\prime}}^2  -\frac{9}{4} y_{\tau}^2 \right)+ \\ 
			\nonumber
			&  ~~~~~~~~ +\,  6 \left(\lambda^2 - 2 y_{\rm t}^4 -2 \lambda y_{\rm t}^2\right)  +  |Y|^2 \left(\frac{5}{8} y_{\rm t}^2-\frac{9}{8} y_{\rm t}^2 - \frac{9}{4} |Y|^2 + \frac{5}{8} {g^\prime}^2 +  \frac{15}{8} g^2 \right) \bigg]  \\
			\nonumber
			&+  \frac{y_{\rm t}}{(4 \pi)^6} \bigg[ y_{\rm t}^4 \left(58.6028 y_{\rm t}^2 + 198 \lambda - 157 g_s^2 -  \frac{1593}{16} g^2 - \frac{2437}{48} {g^\prime}^2\right)  \\
			\nonumber
			&  ~~~~~~~~ +\,\lambda y_{\rm t}^2 \left( \frac{15}{4} \lambda + 16 g_s^2 - \frac{135}{2} g^2 - \frac{127}{6} {g^\prime}^2\right)  \\
			\nonumber
			&  ~~~~~~~~ +\,y_{\rm t}^2 \left(363.764 g_s^4 + 16.990 g^4 - 67.839 {g^\prime}^4 + 48.370 g_s^2 g^2 + 30.123 g_s^2 {g^\prime}^2 + 58.048 g^2 {g^\prime}^2 \right)  \\
			\nonumber
			&  ~~~~~~~~ +\,\lambda^2 \left(-36 \lambda + 45 g^2 + 15 {g^\prime}^2\right) + \lambda \left(- \frac{171}{16} g^4 - \frac{1089}{144} {g^\prime}^4 + \frac{39}{8} g^2 {g^\prime}^2\right) \\
			\nonumber  
			&  ~~~~~~~~ -\,  619.35 g_s^6 + 169.829 g^6 + 74.074 {g^\prime}^6 +  73.654 g_s^4 g^2 - 25.16 g_s^4 {g^\prime}^2  \\
			\nonumber
			&  ~~~~~~~~ -\, 21.072 g_s^2 g^4 - 61.997 g_s^2 {g^\prime}^4 -\frac{107}{4} g_s^2 g^2 {g^\prime}^2 - 7.905 g^4 {g^\prime}^2 - 12.339 g^2 {g^\prime}^4 \bigg]  + \mathcal{O} \left( \frac{m^2_h}{\Lambda^2_f} \right)\,.
		\end{align}
		Meanwhile, the beta function of the bottom Yukawa is, at two-loop order,

		\begin{align}
			\nonumber  \beta_{y_{\rm b}} = \frac{y_{\rm b}}{(4 \pi)^2} \bigg[&  \frac{3}{2} y_{\rm t}^2 + \frac{9}{2}  y_{\rm b}^2 + y_{\tau}^2 - 8 g_s^2 - \frac{9}{4} g^2 - \frac{5}{12} {g^\prime}^2 + |Y|^2) \bigg]+ \\  \nonumber
			+ \frac{y_{\rm b}}{(4 \pi)^4} \bigg[& \frac{5}{8} y_{\rm t}^2 |Y|^2 - \frac{9}{8}  y_{\rm b}^2 |Y|^2 - \frac{9}{4}  |Y|^4 + \frac{15}{8}|Y|^2 ( \frac{1}{3} {g^\prime}^2 + g^2) + 6 \lambda^2 - 108 g_s^4 -  \frac{23}{4} {g^\prime}^4 - \frac{127}{216} {g^\prime}^4 + \\  \nonumber
			&+ 9 g_s^2 g^2 + \frac{31}{9} g_s^2 {g^\prime}^2 - \frac{9}{4} g^2 {g^\prime}^2 +   y_{\rm t}^2 (- \frac{1}{4} y_{\rm t}^2 - \frac{11}{4} y_{\rm b}^2 + \frac{5}{4} y_{\tau}^2 + 4 g_s^2 +  \frac{99}{16} g^2 + \frac{91}{48} {g^\prime}^2) \\  \nonumber
			&++ y_{\rm b}^2 (-12 y_{\rm b}^2 - \frac{9}{4} y_{\tau}^2 - 12 \lambda + 36 g_s^2 + \frac{225}{16} g^2 + \frac{79}{16} {g^\prime}^2)  + y_{\tau}^2 (-\frac{9}{4} y_{\tau}^2 + \frac{15}{8} g^2 + \frac{25}{8} {g^\prime}^2) \bigg].
		\end{align}
		
		At the same order, the tau Yukawa's beta function is 
		
		\begin{align}
			\nonumber   \beta_{y_{\tau}} = \frac{y_{\tau}}{(4 \pi)^2} \bigg[& 3 y_{\rm t}^2 + 3 y_{\rm b}^2 + \frac{5}{2} y_{\tau}^2 - \frac{9}{4} g^2 - \frac{15}{4} {g^\prime}^2 \bigg]+ \\  \nonumber
			+\frac{y_{\tau}}{(4 \pi)^2} \bigg[& 6 \lambda^2 - \frac{23}{4} g^4 + \frac{457}{24} {g^\prime}^4 +   \frac{9}{4} g^2 {g^\prime}^2 +  y_{\rm t}^2 (- \frac{27}{4} y_{\rm t}^2 + \frac{3}{2} y_{\rm b}^2 - \frac{27}{4} y_{\tau}^2 +20 g_s^2 + \frac{45}{8} g^2 + \frac{85}{24} {g^\prime}^2) + \\
			&+ y_{\rm b}^2 (-\frac{27}{4} y_{\rm b}^2 - \frac{27}{4} y_{\tau}^2 + 20 g_s^2 + \frac{45}{8} g^2 +   \frac{25}{24} {g^\prime}^2) +  y_{\tau}^2 (-3 y_{\tau}^2 - 12 \lambda + 165 g^2 + \frac{179}{16} {g^\prime}^2) \bigg].
		\end{align}
		
		The beta functions of the gauge couplings~$g^\prime$,~$g$ and~$g_s$ are, again at three-loop order, respectively given by
		\begin{align}
			\nonumber
			\beta_{g^\prime} =& \frac{{g^{\prime}}^3}{(4 \pi)^2} \frac{41}{6} + \frac{{g^{\prime}}^3}{(4 \pi)^4}\bigg[ \frac{199}{18}{g^{\prime}}^2 + \frac{9}{2}g^2 + \frac{44}{3} g_s^2 - \frac{17}{6} y_{\rm t}^2 - \frac{1}{2} |Y|^2 \bigg]  \\ 
			\nonumber
			&+ \frac{{g^{\prime}}^3}{(4 \pi)^6}\bigg[ y_{\rm t}^2 \bigg( \frac{315}{16} y_{\rm t}^2 -\frac{29}{5} g_s^2 - \frac{785}{32} g^2 - \frac{2827}{288} {g^\prime}^2 \bigg) + \lambda \bigg(-3 \lambda + \frac{3}{2} g^2 + \frac{3}{2} {g^\prime }^2 \bigg)  \\ 
			&  ~~~~~~~~ +\, 99 g_s^4 + \frac{1315}{64} g^4 - \frac{388613}{5184} {g^\prime}^4 - \frac{25}{9} g_s^2 g^2 - \frac{137}{27} g_s^2 {g^\prime}^2 +\frac{205}{96} g^2 {g^\prime}^2 \bigg]   \,; \\
			\nonumber
			\beta_{g} =& - \frac{g^3}{(4 \pi)^2}\frac{19}{6}  + \frac{g^3}{(4 \pi)^4}\bigg[ \frac{3}{2}{g^{\prime}}^2 + \frac{35}{6}g^2 + 12 g_s^2 - \frac{3}{2} y_{\rm t}^2 - \frac{1}{2} |Y|^2 \bigg]  \\ 
			\nonumber
			&+ \frac{g^3}{(4 \pi)^6}\bigg[ y_{\rm t}^2 \left( \frac{147}{16} y_{\rm t}^2 - 7 g_s^2 - \frac{729}{32} g^2 - \frac{593}{96} {g^\prime}^2 \right) + \lambda \left( - 3 \lambda + \frac{3}{2} g^2 + \frac{1}{2} {g^\prime}^2 \right) \\ 
			&  ~~~~~~~~ +\, 81 g_s^4 + \frac{324953}{1728} g^4 - \frac{5597}{576} {g^\prime}^4 + 39 g_s^2 g^2 - \frac{1}{3} g_s^2 {g^\prime}^2 + \frac{291}{32} g^2 {g^\prime}^2 \bigg]
		\end{align}
		and
		\begin{align}
			\nonumber
			\beta_{g_s} =&- \frac{g_s^3}{(4 \pi)^2} 7  + \frac{g_s^3}{(4 \pi)^4}\bigg[ \frac{11}{6}g_s^2 + \frac{9}{2}g^2 - 26 g_s^2 - 2 y_{\rm t}^2 \bigg] \\ 
			\nonumber
			&+  \frac{g_s^3}{(4 \pi)^6} \bigg[y_{\rm t}^2 \left(15 y_{\rm t}^2 - 40 g_s^2 - 93/8 g^2 - 101/24 {g^\prime}^2\right) \\ 
			&  ~~~~~~~~ +\, \frac{65}{2} g_s^4 + \frac{109}{8} g^4 - \frac{2615}{216} {g^\prime}^4 +  21 g_s^2 g^2 + \frac{77}{9} g_s^2 {g^\prime}^2 - \frac{1}{8} g^2 {g^\prime}^2 \bigg]\,.
		\end{align}
		These are at 3-loop order, except for~$\beta_{g_s}$ which includes the dominant 4-loop term. For the neutrinos' Yukawa couplings, the beta function at 2-loop order is 
		\begin{align}
			\nonumber
			\beta_{Y_i} =& \frac{Y_i}{(4 \pi)^2} \bigg[\frac{5}{2} |Y|^2+3 y_{\rm t}^2 - \frac{3}{4} {g^\prime}^2 - \frac{9}{4} g^2 \bigg] \\
			\nonumber
			&+ \frac{Y_i}{(4 \pi)^4} \bigg[ \frac{3}{2} |Y|^4 - \frac{9}{4} |Y|^2 \left(3 y_{\rm t}^2 + |Y|^2\right)   -  \frac{9}{4} \left(3 y_{\rm t}^4 + |Y|^4\right) + \frac{3}{2} \lambda^2 - 64 \lambda |Y|^2 + \frac{|Y|^2}{16} \left(93 {g^\prime}^2 + 135 g^2\right)  \\ 
			\nonumber
			&  ~~~~~~~~ +\,  \frac{5}{2} \left(y_{\rm t}^2 \left(\frac{17}{12} {g^\prime}^2 + \frac{9}{4} g^2 + 8 g_s^2\right) + \frac{3}{4} |Y|^2 \left(\frac{1}{3} {g^\prime}^2 + g^2\right)\right) + \frac{7}{48} {g^\prime}^4 - \frac{9}{4} g^2 {g^\prime}^2 - \frac{23}{4} g^4\bigg] + \mathcal{O} \left( \frac{m^2_h}{\Lambda^2_f} \right) \,. \\
		\end{align}
		The running of the Wilson coefficient~$C_6$ at 1-loop order is determined by 
		\begin{equation}
			\beta_{C_6}= \frac{C_6}{(4 \pi)^2} \left[- \frac{9}{2} \left(3 g^2 + {g^\prime}^2\right) + 108 \lambda + 18 y_{\rm t}^2 \right]\,.
		\end{equation}
		Lastly, the running mass~$m_h^2$ satisfies
		\begin{equation}
			\beta_{m_h^2 }= \frac{3 m_h^2}{8 \pi^2}\left[ 2 \lambda +y_{\rm t}^2 - \frac{3}{4}g^2 - \frac{3}{20}{g^\prime}^2 + \mathcal{O} \left( \frac{m^2_h}{\Lambda^2_f} \right)\right]\,.
		\end{equation}
		
	\end{widetext}


\begin{thebibliography}{9}
		
		\bibitem{Giudice:2017pzm} 
		G.~F.~Giudice,
		``The Dawn of the Post-Naturalness Era,''
		arXiv:1710.07663 [physics.hist-ph].
		
		\bibitem{tHooft:1979rat}
		G.~'t Hooft,
		``Naturalness, chiral symmetry, and spontaneous chiral symmetry breaking,''
		NATO Sci. Ser. B \textbf{59}, 135-157 (1980)
		
		\bibitem{Frampton:1976kf} 
		P.~H.~Frampton,
		``Vacuum Instability and Higgs Scalar Mass,''
		Phys.\ Rev.\ Lett.\  {\bf 37}, 1378 (1976)
		Erratum: [Phys.\ Rev.\ Lett.\  {\bf 37}, 1716 (1976)].
		
		\bibitem{Sher:1988mj} 
		M.~Sher,
		``Electroweak Higgs Potentials and Vacuum Stability,''
		Phys.\ Rept.\  {\bf 179}, 273 (1989).
		
		\bibitem{Casas:1994qy} 
		J.~A.~Casas, J.~R.~Espinosa and M.~Quiros,
		``Improved Higgs mass stability bound in the standard model and implications for supersymmetry,''
		Phys.\ Lett.\ B {\bf 342}, 171 (1995)
		[hep-ph/9409458].
		
		\bibitem{Espinosa:1995se} 
		J.~R.~Espinosa and M.~Quiros,
		``Improved metastability bounds on the standard model Higgs mass,''
		Phys.\ Lett.\ B {\bf 353}, 257 (1995)
		[hep-ph/9504241].
		
		\bibitem{Isidori:2001bm} 
		G.~Isidori, G.~Ridolfi and A.~Strumia,
		``On the metastability of the standard model vacuum,''
		Nucl.\ Phys.\ B {\bf 609}, 387 (2001)
		[hep-ph/0104016].
		
		\bibitem{Espinosa:2007qp} 
		J.~R.~Espinosa, G.~F.~Giudice and A.~Riotto,
		``Cosmological implications of the Higgs mass measurement,''
		JCAP {\bf 0805}, 002 (2008)
		[arXiv:0710.2484 [hep-ph]].
		
		\bibitem{Ellis:2009tp} 
		J.~Ellis, J.~R.~Espinosa, G.~F.~Giudice, A.~Hoecker and A.~Riotto,
		``The Probable Fate of the Standard Model,''
		Phys.\ Lett.\ B {\bf 679}, 369 (2009)
		[arXiv:0906.0954 [hep-ph]].
		
		\bibitem{Degrassi:2012ry} 
		G.~Degrassi, S.~Di Vita, J.~Elias-Miro, J.~R.~Espinosa, G.~F.~Giudice, G.~Isidori and A.~Strumia,
		``Higgs mass and vacuum stability in the Standard Model at NNLO,''
		JHEP {\bf 1208}, 098 (2012)
		[arXiv:1205.6497 [hep-ph]].
		
		\bibitem{Buttazzo:2013uya} 
		D.~Buttazzo, G.~Degrassi, P.~P.~Giardino, G.~F.~Giudice, F.~Sala, A.~Salvio and A.~Strumia,
		``Investigating the near-criticality of the Higgs boson,''
		JHEP {\bf 1312}, 089 (2013)
		[arXiv:1307.3536 [hep-ph]].
		
		\bibitem{Lalak:2014qua} 
		Z.~Lalak, M.~Lewicki and P.~Olszewski,
		``Higher-order scalar interactions and SM vacuum stability,''
		JHEP {\bf 1405}, 119 (2014)
		[arXiv:1402.3826 [hep-ph]].
		
		\bibitem{Andreassen:2014gha} 
		A.~Andreassen, W.~Frost and M.~D.~Schwartz,
		``Consistent Use of the Standard Model Effective Potential,''
		Phys.\ Rev.\ Lett.\  {\bf 113}, no. 24, 241801 (2014)
		[arXiv:1408.0292 [hep-ph]].
		
		\bibitem{Branchina:2014rva} 
		V.~Branchina, E.~Messina and M.~Sher,
		``Lifetime of the electroweak vacuum and sensitivity to Planck scale physics,''
		Phys.\ Rev.\ D {\bf 91}, 013003 (2015)
		[arXiv:1408.5302 [hep-ph]].
		
		\bibitem{Bednyakov:2015sca} 
		A.~V.~Bednyakov, B.~A.~Kniehl, A.~F.~Pikelner and O.~L.~Veretin,
		``Stability of the Electroweak Vacuum: Gauge Independence and Advanced Precision,''
		Phys.\ Rev.\ Lett.\  {\bf 115}, no. 20, 201802 (2015)
		[arXiv:1507.08833 [hep-ph]].
		
		\bibitem{Iacobellis:2016eof} 
		G.~Iacobellis and I.~Masina,
		``Stationary configurations of the Standard Model Higgs potential: electroweak stability and rising inflection point,''
		Phys.\ Rev.\ D {\bf 94}, no. 7, 073005 (2016)
		[arXiv:1604.06046 [hep-ph]].
		
		\bibitem{AFS}
		A.~Andreassen, W.~Frost and M.~D.~Schwartz,
		``Scale Invariant Instantons and the Complete Lifetime of the Standard Model,''
		Phys.\ Rev.\ D {\bf 97}, no. 5, 056006 (2018)
		[arXiv:1707.08124 [hep-ph]].
		
		\bibitem{Zyla:2020zbs}
		P.A.~Zyla \textit{et al.} [Particle Data Group],
		``Review of Particle Physics,''
		PTEP \textbf{2020}, no.8, 083C01 (2020).
		
		\bibitem{Huang:2020hdv}
		G.~y.~Huang and S.~Zhou,
		``Precise Values of Running Quark and Lepton Masses in the Standard Model,''
		Phys. Rev. D \textbf{103}, no.1, 016010 (2021)
		[arXiv:2009.04851 [hep-ph]].
		
		%
		%
		
		\bibitem{Guth:1982pn} 
		A.~H.~Guth and E.~J.~Weinberg,
		``Could the Universe Have Recovered from a Slow First Order Phase Transition?,''
		Nucl.\ Phys.\ B {\bf 212}, 321 (1983).
		
		\bibitem{Giudice:2006sn} 
		G.~F.~Giudice and R.~Rattazzi,
		``Living Dangerously with Low-Energy Supersymmetry,''
		Nucl.\ Phys.\ B {\bf 757}, 19 (2006)
		[hep-ph/0606105].
		
		\bibitem{Samuel:1999am} 
		S.~Samuel,
		``The Standard model in its other phase,''
		Nucl.\ Phys.\ B {\bf 597}, 70 (2001)
		[hep-ph/9910559].
		
		\bibitem{ArkaniHamed:2005yv} 
		N.~Arkani-Hamed, S.~Dimopoulos and S.~Kachru,
		``Predictive landscapes and new physics at a TeV,''
		hep-th/0501082.
		
		
		
		\bibitem{Steinhardt:1982kg}
		P.~J.~Steinhardt,
		``Natural inflation,''
		Contribution to the ``Nuffield Workshop on the Very Early Universe", p. 251,
		UPR-0198T.
		
		\bibitem{Vilenkin:1983xq}
		A.~Vilenkin,
		``The Birth of Inflationary Universes,''
		Phys. Rev. D \textbf{27}, 2848 (1983).
		
		\bibitem{Linde:1986fc}
		A.~D.~Linde,
		``Eternal chaotic inflation,"
		Mod. Phys. Lett. A \textbf{1}, 81 (1986).
		
		\bibitem{Linde:1986fd}
		A.~D.~Linde,
		``Eternally Existing Selfreproducing Chaotic Inflationary Universe,''
		Phys. Lett. B \textbf{175}, 395-400 (1986).
		
		\bibitem{Starobinsky:1986fx}
		A.~A.~Starobinsky,
		``Stochastic de Sitter (inflationary) stage in the early universe,"
		Lect. Notes Phys. \textbf{246}, 107-126 (1986). 
		
		%
		
		
		\bibitem{Garriga:2005av} 
		J.~Garriga, D.~Schwartz-Perlov, A.~Vilenkin and S.~Winitzki,
		``Probabilities in the inflationary multiverse,''
		JCAP {\bf 0601}, 017 (2006)
		[hep-th/0509184].
		
		\bibitem{Denef:2017cxt} 
		F.~Denef, M.~R.~Douglas, B.~Greene and C.~Zukowski,
		``Computational complexity of the landscape II - Cosmological considerations,''
		Annals Phys.\  {\bf 392}, 93 (2018)
		[arXiv:1706.06430 [hep-th]].
		
		\bibitem{Khoury:2019yoo}
		J.~Khoury and O.~Parrikar,
		``Search Optimization, Funnel Topography, and Dynamical Criticality on the String Landscape,''
		JCAP \textbf{12}, 014 (2019)
		[arXiv:1907.07693 [hep-th]].
		
		\bibitem{Khoury:2019ajl}
		J.~Khoury,
		``Accessibility Measure for Eternal Inflation: Dynamical Criticality and Higgs Metastability,''
		JCAP \textbf{06}, 009 (2021)
		[arXiv:1912.06706 [hep-th]].
		
		\bibitem{Kartvelishvili:2020thd}
		G.~Kartvelishvili, J.~Khoury and A.~Sharma,
		``The Self-Organized Critical Multiverse,''
		JCAP \textbf{02}, 028 (2021)
		[arXiv:2003.12594 [hep-th]].
		
		\bibitem{Khoury:2021grg}
		J.~Khoury and S.~S.~C.~Wong,
		``Early-Time Measure in Eternal Inflation,''
		[arXiv:2106.12590 [hep-th]].
		
		\bibitem{QBT}
		G.~Dvali,
		``$S$-Matrix and Anomaly of de Sitter,''
		Symmetry \textbf{13}, no.1, 3 (2020)
		[arXiv:2012.02133 [hep-th]].
		
		
		\bibitem{Geller:2018xvz}
		M.~Geller, Y.~Hochberg and E.~Kuflik,
		``Inflating to the Weak Scale,''
		Phys. Rev. Lett. \textbf{122}, no.19, 191802 (2019)
		doi:10.1103/PhysRevLett.122.191802
		[arXiv:1809.07338 [hep-ph]].
		
		\bibitem{Giudice:2021viw}
		G.~F.~Giudice, M.~McCullough and T.~You,
		``Self-Organised Localisation,''
		[arXiv:2105.08617 [hep-ph]].
		
		\bibitem{Bardeen:1995kv}
		W.~A.~Bardeen,
		``On naturalness in the standard model,''
		FERMILAB-CONF-95-391-T.
		
		\bibitem{Hempfling:1996ht}
		R.~Hempfling,
		``The Next-to-minimal Coleman-Weinberg model,''
		Phys. Lett. B \textbf{379}, 153-158 (1996)
		[arXiv:hep-ph/9604278 [hep-ph]].
		
		\bibitem{Meissner:2006zh}
		K.~A.~Meissner and H.~Nicolai,
		``Conformal Symmetry and the Standard Model,''
		Phys. Lett. B \textbf{648}, 312-317 (2007)
		[arXiv:hep-th/0612165 [hep-th]].
		
		\bibitem{HierCrit}
		G.~F.~Giudice and R.~Rattazzi,
		``Living Dangerously with Low-Energy Supersymmetry,''
		Nucl. Phys. B \textbf{757}, 19-46 (2006)
		[arXiv:hep-ph/0606105 [hep-ph]].
		
		\bibitem{B-L}	                
		M.~Shaposhnikov,
		``A Possible symmetry of the nuMSM,''
		Nucl. Phys. B \textbf{763}, 49-59 (2007)
		[arXiv:hep-ph/0605047 [hep-ph]].
		
		\bibitem{nufeat1}	               
		F.~Bezrukov, M.~Y.~Kalmykov, B.~A.~Kniehl and M.~Shaposhnikov,
		``Higgs Boson Mass and New Physics,''
		JHEP \textbf{10}, 140 (2012)
		[arXiv:1205.2893 [hep-ph]].
		
		\bibitem{nufeat2}	                
		T.~Asaka and M.~Shaposhnikov,
		``The $\nu$MSM, dark matter and baryon asymmetry of the universe,''
		Phys. Lett. B \textbf{620}, 17-26 (2005)
		[arXiv:hep-ph/0505013 [hep-ph]].
		
		\bibitem{Asaka:2005an}
		T.~Asaka, S.~Blanchet and M.~Shaposhnikov,
		``The nuMSM, dark matter and neutrino masses,''
		Phys. Lett. B \textbf{631}, 151-156 (2005)
		[arXiv:hep-ph/0503065 [hep-ph]].
		
		
		\bibitem{nufeat3}	                
		M.~Shaposhnikov,
		``The nuMSM, leptonic asymmetries, and properties of singlet fermions,''
		JHEP \textbf{08}, 008 (2008)
		[arXiv:0804.4542 [hep-ph]].
		
		\bibitem{nufeat4}
		L.~Canetti, M.~Drewes, T.~Frossard and M.~Shaposhnikov,
		``Dark Matter, Baryogenesis and Neutrino Oscillations from Right Handed Neutrinos,''
		Phys. Rev. D \textbf{87}, 093006 (2013)
		[arXiv:1208.4607 [hep-ph]].
		
		\bibitem{nufeat5}	                
		J.~Ghiglieri and M.~Laine,
		``Sterile neutrino dark matter via coinciding resonances,''
		JCAP \textbf{07}, 012 (2020)
		[arXiv:2004.10766 [hep-ph]].
		
		\bibitem{Coleman}
		S.~R.~Coleman,
		``The Fate of the False Vacuum. 1. Semiclassical Theory,''
		Phys.\ Rev.\ D {\bf 15}, 2929 (1977)
		Erratum: [Phys.\ Rev.\ D {\bf 16}, 1248 (1977)].
		
		\bibitem{Callan:1977pt} 
		C.~G.~Callan, Jr. and S.~R.~Coleman,
		``The Fate of the False Vacuum. 2. First Quantum Corrections,''
		Phys.\ Rev.\ D {\bf 16}, 1762 (1977).
		
		\bibitem{Coleman:1980aw} 
		S.~R.~Coleman and F.~De Luccia,
		``Gravitational Effects on and of Vacuum Decay,''
		Phys.\ Rev.\ D {\bf 21}, 3305 (1980).
		
		\bibitem{Fubini}
		S.~Fubini,
		``A New Approach to Conformal Invariant Field Theories,''
		Nuovo Cim. A \textbf{34}, 521 (1976).
		
		\bibitem{Lipatov}
		L.~N.~Lipatov,
		``Divergence of the Perturbation Theory Series and the Quasiclassical Theory,''
		Sov. Phys. JETP \textbf{45}, 216-223 (1977).
		
		\bibitem{Jose}
		J.~R.~Espinosa,
		``Vacuum Decay in the Standard Model: Analytical Results with Running and Gravity,''
		JCAP \textbf{06}, 052 (2020)
		[arXiv:2003.06219 [hep-ph]].
		
		\bibitem{Gra}	            
		G.~Isidori, V.~S.~Rychkov, A.~Strumia and N.~Tetradis,
		``Gravitational corrections to standard model vacuum decay,''
		Phys. Rev. D \textbf{77}, 025034 (2008)
		[arXiv:0712.0242 [hep-ph]].
		
		\bibitem{Grav}		         
		A.~Salvio, A.~Strumia, N.~Tetradis and A.~Urbano,
		``On gravitational and thermal corrections to vacuum decay,''
		JHEP \textbf{09}, 054 (2016)
		[arXiv:1608.02555 [hep-ph]].
		
		\bibitem{Consistent}                		
		A.~Andreassen, W.~Frost and M.~D.~Schwartz,
		``Consistent Use of the Standard Model Effective Potential,''
		Phys.\ Rev.\ Lett.\  {\bf 113}, no. 24, 241801 (2014)
		[arXiv:1408.0292 [hep-ph]].
		
		\bibitem{Gravcorcor}	            
		A.~Rajantie and S.~Stopyra,
		``Standard Model vacuum decay with gravity,''
		Phys. Rev. D \textbf{95}, no.2, 025008 (2017)
		[arXiv:1606.00849 [hep-th]].
		
		\bibitem{B-LMarko}
		P.~Agrawal, M.~Bauer, J.~Beacham, A.~Berlin, A.~Boyarsky, S.~Cebrian, X.~Cid-Vidal, D.~d'Enterria, A.~De Roeck and M.~Drewes, \textit{et al.}
		``Feebly-Interacting Particles:FIPs 2020 Workshop Report,''
		[arXiv:2102.12143 [hep-ph]].
		
		\bibitem{nudecay&threshold}	                
		S.~Chigusa, T.~Moroi and Y.~Shoji,
		``Decay Rate of Electroweak Vacuum in the Standard Model and Beyond,''
		Phys. Rev. D \textbf{97}, no.11, 116012 (2018)
		[arXiv:1803.03902 [hep-ph]].
		
		\bibitem{nudecay1}	               
		I.~Garg, S.~Goswami, K.~N.~Vishnudath and N.~Khan,
		``Electroweak vacuum stability in presence of singlet scalar dark matter in TeV scale seesaw models,''
		Phys. Rev. D \textbf{96}, no.5, 055020 (2017)
		[arXiv:1706.08851 [hep-ph]].
		
		\bibitem{nudecay2}	               
		S.~Khan, S.~Goswami and S.~Roy,
		``Vacuum Stability constraints on the minimal singlet TeV Seesaw Model,''
		Phys. Rev. D \textbf{89}, no.7, 073021 (2014)
		[arXiv:1212.3694 [hep-ph]].
		
		\bibitem{nudecay3}	                
		L.~Delle Rose, C.~Marzo and A.~Urbano,
		``On the stability of the electroweak vacuum in the presence of low-scale seesaw models,''
		JHEP \textbf{12}, 050 (2015)
		[arXiv:1506.03360 [hep-ph]].
		
		\bibitem{nudecay4}	                
		S.~Mandal, R.~Srivastava and J.~W.~F.~Valle,
		``Consistency of the dynamical high-scale type-I seesaw mechanism,''
		Phys. Rev. D \textbf{101}, no.11, 115030 (2020)
		[arXiv:1903.03631 [hep-ph]].
		
		\bibitem{nuSMmatching}                       
		J.~A.~Casas, V.~Di Clemente, A.~Ibarra and M.~Quiros,
		``Massive neutrinos and the Higgs mass window,''
		Phys. Rev. D \textbf{62}, 053005 (2000)
		[arXiv:hep-ph/9904295 [hep-ph]].
		
		\bibitem{nuSMBeta2}                        
		Y.~F.~Pirogov and O.~V.~Zenin,
		``Two loop renormalization group restrictions on the standard model and the fourth chiral family,''
		Eur. Phys. J. C \textbf{10}, 629-638 (1999)
		[arXiv:hep-ph/9808396 [hep-ph]].
		
		\bibitem{nuEFT}                      
		S.~Antusch, J.~Kersten, M.~Lindner and M.~Ratz,
		``Neutrino mass matrix running for nondegenerate seesaw scales,''
		Phys. Lett. B \textbf{538}, 87-95 (2002)
		[arXiv:hep-ph/0203233 [hep-ph]].
		
		\bibitem{nuEFT2}                     
		S.~Antusch, J.~Kersten, M.~Lindner, M.~Ratz and M.~A.~Schmidt,
		``Running neutrino mass parameters in see-saw scenarios,''
		JHEP \textbf{03}, 024 (2005)
		[arXiv:hep-ph/0501272 [hep-ph]].
		
		\bibitem{NExp}                    
		M.~Chrzaszcz, M.~Drewes, T.~E.~Gonzalo, J.~Harz, S.~Krishnamurthy and C.~Weniger,
		``A frequentist analysis of three right-handed neutrinos with GAMBIT,''
		Eur. Phys. J. C \textbf{80}, no.6, 569 (2020)
		[arXiv:1908.02302 [hep-ph]].
		
		\bibitem{CHPotreview}	                
		G.~Cacciapaglia and F.~Sannino,
		``Fundamental Composite (Goldstone) Higgs Dynamics,''
		JHEP \textbf{04}, 111 (2014)
		[arXiv:1402.0233 [hep-ph]].
		
		\bibitem{thetatuning}
		G.~Cacciapaglia, C.~Pica and F.~Sannino,
		``Fundamental Composite Dynamics: A Review,''
		Phys. Rept. \textbf{877}, 1-70 (2020)
		[arXiv:2002.04914 [hep-ph]].
		
		\bibitem{thetatuningweak}
		D.~Buarque Franzosi, G.~Cacciapaglia and A.~Deandrea,
		``Sigma-assisted low scale composite Goldstone\textendash{}Higgs,''
		Eur. Phys. J. C \textbf{80}, no.1, 28 (2020)
		[arXiv:1809.09146 [hep-ph]].
		
		\bibitem{SMEFTBeta}		    
		A.~Celis, J.~Fuentes-Martin, A.~Vicente and J.~Virto,
		``DsixTools: The Standard Model Effective Field Theory Toolkit,''
		Eur. Phys. J. C \textbf{77}, no.6, 405 (2017)
		[arXiv:1704.04504 [hep-ph]].
		
		
		
		\bibitem{PCHiggs}
		T.~Alanne, D.~Buarque Franzosi, M.~T.~Frandsen, M.~L.~A.~Kristensen, A.~Meroni and M.~Rosenlyst,
		``Partially composite Higgs models: Phenomenology and RG analysis,''
		JHEP \textbf{01}, 051 (2018)
		[arXiv:1711.10410 [hep-ph]].
		
		\bibitem{Chang:2007ki}
		W.~F.~Chang, J.~N.~Ng and J.~M.~S.~Wu,
		``Shadow Higgs from a scale-invariant hidden U(1)(s) model,''
		Phys. Rev. D \textbf{75}, 115016 (2007)
		[arXiv:hep-ph/0701254 [hep-ph]].
		
		\bibitem{Foot:2007as}
		R.~Foot, A.~Kobakhidze and R.~R.~Volkas,
		``Electroweak Higgs as a pseudo-Goldstone boson of broken scale invariance,''
		Phys. Lett. B \textbf{655}, 156-161 (2007)
		[arXiv:0704.1165 [hep-ph]].
		
		\bibitem{Foot:2007iy}
		R.~Foot, A.~Kobakhidze, K.~L.~McDonald and R.~R.~Volkas,
		``A Solution to the hierarchy problem from an almost decoupled hidden sector within a classically scale invariant theory,''
		Phys. Rev. D \textbf{77}, 035006 (2008)
		[arXiv:0709.2750 [hep-ph]].
		
		\bibitem{Iso:2009ss}
		S.~Iso, N.~Okada and Y.~Orikasa,
		``Classically conformal $B^-$ L extended Standard Model,''
		Phys. Lett. B \textbf{676}, 81-87 (2009)
		[arXiv:0902.4050 [hep-ph]].
		
		\bibitem{Alexander-Nunneley:2010tyr}
		L.~Alexander-Nunneley and A.~Pilaftsis,
		``The Minimal Scale Invariant Extension of the Standard Model,''
		JHEP \textbf{09}, 021 (2010)
		[arXiv:1006.5916 [hep-ph]].
		
		\bibitem{Iso:2012jn}
		S.~Iso and Y.~Orikasa,
		``TeV Scale B-L model with a flat Higgs potential at the Planck scale: In view of the hierarchy problem,''
		PTEP \textbf{2013}, 023B08 (2013)
		[arXiv:1210.2848 [hep-ph]].
		
		\bibitem{Englert:2013gz}
		C.~Englert, J.~Jaeckel, V.~V.~Khoze and M.~Spannowsky,
		``Emergence of the Electroweak Scale through the Higgs Portal,''
		JHEP \textbf{04}, 060 (2013)
		[arXiv:1301.4224 [hep-ph]].
		
		\bibitem{Chun:2013soa}
		E.~J.~Chun, S.~Jung and H.~M.~Lee,
		``Radiative generation of the Higgs potential,''
		Phys. Lett. B \textbf{725}, 158-163 (2013)
		[erratum: Phys. Lett. B \textbf{730}, 357-359 (2014)]
		[arXiv:1304.5815 [hep-ph]].
		
		\bibitem{Heikinheimo:2013fta}
		M.~Heikinheimo, A.~Racioppi, M.~Raidal, C.~Spethmann and K.~Tuominen,
		``Physical Naturalness and Dynamical Breaking of Classical Scale Invariance,''
		Mod. Phys. Lett. A \textbf{29}, 1450077 (2014)
		[arXiv:1304.7006 [hep-ph]].
		
		\bibitem{Hambye:2013dgv}
		T.~Hambye and A.~Strumia,
		``Dynamical generation of the weak and Dark Matter scale,''
		Phys. Rev. D \textbf{88}, 055022 (2013)
		[arXiv:1306.2329 [hep-ph]].
		
		\bibitem{Khoze:2013oga}
		V.~V.~Khoze and G.~Ro,
		``Leptogenesis and Neutrino Oscillations in the Classically Conformal Standard Model with the Higgs Portal,''
		JHEP \textbf{10}, 075 (2013)
		[arXiv:1307.3764 [hep-ph]].
		
		\bibitem{Carone:2013wla}
		C.~D.~Carone and R.~Ramos,
		``Classical scale-invariance, the electroweak scale and vector dark matter,''
		Phys. Rev. D \textbf{88}, 055020 (2013)
		[arXiv:1307.8428 [hep-ph]].
		
		\bibitem{Khoze:2013uia}
		V.~V.~Khoze,
		``Inflation and Dark Matter in the Higgs Portal of Classically Scale Invariant Standard Model,''
		JHEP \textbf{11}, 215 (2013)
		[arXiv:1308.6338 [hep-ph]].
		
		\bibitem{Steele:2013fka}
		T.~G.~Steele, Z.~W.~Wang, D.~Contreras and R.~B.~Mann,
		Phys. Rev. Lett. \textbf{112}, no.17, 171602 (2014)
		doi:10.1103/PhysRevLett.112.171602
		[arXiv:1310.1960 [hep-ph]].
		
		\bibitem{Wang:2015cda}
		Z.~W.~Wang, T.~G.~Steele, T.~Hanif and R.~B.~Mann,
		JHEP \textbf{08}, 065 (2016)
		doi:10.1007/JHEP08(2016)065
		[arXiv:1510.04321 [hep-ph]].
		
		
		
		
		
		%
		
		
		
		
		%
		%
		
		
		
		
		
		
		
		
		
		
		
		
		
		
		
		
		
		
		
		
		%
		%
		
		
		
		
		
		
		
		
		
		
		
		
		
		
		%
		
		
		
		
		
		
		
		
	\end{thebibliography}
\end{document}